\documentclass[fleqn,usenatbib]{mnras}

\usepackage{newtxtext,newtxmath}

\usepackage[T1]{fontenc}

\DeclareRobustCommand{\VAN}[3]{#2}
\let\VANthebibliography\thebibliography
\def\thebibliography{\DeclareRobustCommand{\VAN}[3]{##3}\VANthebibliography}

\usepackage{graphicx}	
\usepackage{amsmath}	
\usepackage{amssymb}	

\usepackage{enumerate}
\usepackage[version=3]{mhchem} 
\usepackage{subfig}

\usepackage{amsfonts}
\usepackage{natbib}

\title[Synthetic IR Spectra of Protoplanetary Discs]{Spatially resolving the chemical composition of the planet building blocks \thanks{Based on observations collected at the European Southern Observatory, Chile (ESO IDs : 090.C-0110).}}

\author[A. Matter et al.]{A.~Matter$^{1}$\thanks{E-mail: Alexis.Matter@oca.eu}, F. C. ~Pignatale$^{2,3}$\thanks{E-mail: pignatale@ipgp.fr}, B.~Lopez$^{1}$\\
$^{1}$Universit\'e C\^ote d'Azur, Observatoire de la C\^ote d'Azur, CNRS, Laboratoire Lagrange, France\\
$^{2}$Institut de Physique du Globe de Paris (IPGP), Univ Paris Diderot, CNRS, 1 rue Jussieu, 75005, Paris, France\\
$^{3}$Museum National d'Histoire Naturelle, UMR 7590, CP52, 57 rue Cuvier, Paris, France}

\date{Accepted XXX. Received YYY; in original form ZZZ}

\pubyear{2020}

\begin{document}
\label{firstpage}
\pagerange{\pageref{firstpage}--\pageref{lastpage}}
\maketitle

\begin{abstract}

The inner regions of protoplanetary discs (from $\sim$ 0.1 to 10~au) are the expected birthplace of planets, especially telluric. In those high temperature regions, solids can experience cyclical annealing, vaporisation and recondensation. Hot and warm dusty grains emits mostly in the infrared domain, notably in N-band (8 to 13~$\mu$m). Studying their fine chemistry through mid-infrared spectro-interferometry with the new VLTI instrument MATISSE, which can spatially resolve these regions, requires detailed dust chemistry models. Using radiative transfer, we derived infrared spectra of a fiducial static protoplanetary disc model with different inner disc ($< 1$~au) dust compositions. The latter were derived from condensation sequences computed at LTE for three initial $C/O$ ratios: subsolar ($C/O=0.4$), solar ($C/O=0.54$), and supersolar ($C/O=1$). The three scenarios return very different N-band spectra, especially when considering the presence of sub-micron-sized dust grains.  MATISSE should be able to detect these differences and trace the associated sub-au-scale radial changes. We propose a first interpretation of N-band `inner-disc' spectra obtained with the former VLTI instrument MIDI on three Herbig stars (HD142527, HD144432, HD163296) and one T Tauri star (AS209). Notably, we could associate a supersolar (`carbon-rich') composition for HD142527 and a subsolar (`oxygen-rich') one for HD1444432.  We show that the inner disc mineralogy can be very specific and not related to the dust composition derived from spatially unresolved mid-infrared spectroscopy. We highlight the need for including more complex chemistry when interpreting solid-state spectroscopic observations of the inner regions of discs, and for considering dynamical aspects for future studies.

\end{abstract}




\section{Introduction}
\label{intro}

The spectra of protoplanetary discs are quite complex since they are produced from a combination of emission/absorption bands related to the solid-state features of dust, on top of a continuum. The wavelength of the dust continuum emission peak is associated with the dust temperature, which is related, at first order, to the location of the dust emitting region in the disc \citep{2007ApJ...659..680K}. Therefore, fitting the observed spectra across a wide range of wavelengths is essential to understand the contribution and the composition of dust in different disc regions. 
The inner regions of protoplanetary discs (from about 0.1 to 10~au) are of particular interest since they are the expected birthplace of planets, especially telluric. Emitting mostly in the infrared domain, the hot and warm dust constitute the building blocks of these planets. Deriving their properties and chemistry is essential to infer the nature and composition of the future planets. \\
The MIR domain, especially the {\itshape N} band (8 to 13~$\mu$m), is rich in solid-state features notably produced by the warm amorphous and cristalline silicate dust \citep{2010ARA&A..48...21H}. Therefore, it is a spectral domain relevant to study the fine chemistry of solids and their evolution in the inner disc regions. 
The dust properties (grain size, shape, composition) dramatically affect the resulting emissivity/absorbance of the medium, and then the emission spectra of protoplanetary discs. 
Two steps are generally required to determine, from the observations, the chemical nature and properties of these building blocks: (i) defining a theoretical dust mixture which can match the observed trend and (ii) computing and minimizing the deviations between the observed and theoretical emission profiles \citep[e.g., ][]{2005A&A...437..189V,2010ApJ...721..431J}.\\ 
Mid-infrared (3~${\mu}$m to 180~${\mu}$m) spectroscopic observations with the Infrared Space Observatory (ISO) or the {\itshape Spitzer Space Telescope} provided a large database of spatially unresolved IR spectra of protoplanetary discs. Coupled with theoretical models, such observations indicated that amorphous pyroxenes and olivines, crystalline forsterite (\ce{Mg2SiO4}) and enstatite (\ce{MgSiO3}), and amorphous silica (\ce{SiO2}) are ubiquitous in the surface layer of discs \citep{1977ApJ...217..425M,Grossman1972,Gail1998,2005A&A...437..189V,2010ApJ...721..431J}. Thus, it is generally assumed that the majority of the dust in discs is composed of these chemical species. Parallel studies have added the contribution of polycyclic aromatic hydrocarbons to the MIR spectra \citep{2001ApJ...551..807D,2007ApJ...657..810D,2008ApJ...684..411K}. 

 Radial variations of the discs mineralogy could be first addressed through the modelling of the 10~$\mu$m and 20-30~$\mu$m silicates features, assuming that the 10~$\mu$m one is representative of the inner warmer disc regions while the 20-30~$\mu$m one characterizes the outer cooler disc  \citep[]{2006ApJ...639..275K,2008ApJ...683..479B,2009A&A...497..379M}. Infrared observations of T-Tauri disc's upper layers suggested radial variations of the forsterite and enstatite distribution, with more enstatite in the warm inner regions than in the cooler outer regions where the contribution of forsterite would be stronger \citep[]{2006ApJ...639..275K,2008ApJ...683..479B,2009A&A...497..379M}. In parallel, MIR interferometric observations of three Herbig Ae stars by \citet{2004Natur.432..479V} resolved the 1-2~au inner disc region. \citet{2004Natur.432..479V} derived a dominant fractional abundance of crystalline forsterite, which is needed to best fit the 10~$\mu$m region of the observed spectra. This finding of a forsterite-rich dust in the innermost hotter disc zone is compatible with the high temperature condensation of forsterite \citep{2004A&A...413..571G}.

 The high temperatures found in the disc regions close to the protostar would imply that a single dust mixture for the whole disc radial extent would not  predict accurately the real dust components. Indeed, those high temperatures would allow the annealing, vaporisation and recondensation of the dust that will thus change into several phases. As a consequence, we do expect a sensitive radial change in the dust composition. \citet{2016MNRAS.457.1359P} derived 2D-map of condensates distribution within 1~au from the protosun using a 1~Myr old static disc from \citet{Dalessio1998,1999ApJ...527..893D} and solar abundances. They obtained two radial zones at the surface of the inner disc: a "high-temperature" ($T>1000$~K) zone for which the forsterite to enstatite ratio is $fo/en< 1$, and a low-temperature ($500<T\rm{(K)}<1000$) zone with $fo/en > 1$. Nevertheless, when considering the global contribution of the different dust species within the whole 1 au disc surface, forsterite is still the dominant silicate species \citep{2016MNRAS.457.1359P}. These results are in agreement with disc observations \citep[]{2006ApJ...639..275K,2008ApJ...683..479B,2009A&A...497..379M}, which suggests that high temperature equilibrium calculations can represent a valid method to derive the dust composition of the hot inner disc regions. It is especially relevant when considering different bulk elemental compositions of the disc, which are expected to return widely distinctive condensates compositions. 
 For instance, when solar $C/O$ and $Mg/Si$ ratios are considered for the bulk disc composition, thermodynamical equilibrium calculations in the high temperature regions show no presence of \ce{SiO2} and graphite. Indeed, in that case, carbon is bound in \ce{CO}(g) \citep{1995GeCoA..59.3413Y,Gail1998}, which makes graphite not a stable phase, while it is usually considered as a major carrier of the dust continuum emission.
 Varying the $C/O$ ratio would thus result in the appearance/disappearance, to various degrees, of oxides, silicates, and carbon-binding solid compounds \citep{1975GeCoA..39..389L,1999IAUS..191..279L,2012ApJ...757..192J}. As mentioned before, thermodynamics calculations using a solar bulk composition predict that most of the carbon would be locked in CO(g) \citep{2016MNRAS.457.1359P}. Different $C/O$ ratios would thus result in different observable dust mixtures in protoplanetary discs, and eventually in different planetary systems. 
 Measurements of the elemental $C/O$ ratio in different samples of nearby solar-type (F, G, and K) stars indicate a rather wide spread around the solar value ($C/0=0.54$). For instance, \citet{2010ApJ...725.2349D} derived an interval of $0.2<C/O<1.2$ from the HARPS GTO sample, while more recent measurements based for instance on an extended HARPS sample, narrowed it down to $0.2<C/O<0.8$ \citep{2014A&A...568A..25N,2016ApJ...831...20B,2018A&A...614A..84S}. Such a spread in stellar $C/O$ ratio should imply differences in the resulting dust composition of resolved inner discs regions, if star and disc form from the same initial gas and dust reservoir. Those latter observations possibly highlight an interesting dichotomy between planet-hosting stars that have different chemistry and discs that show, instead, an average similar dust composition as mentioned before. If this dichotomy is the consequence of some observational/detection/modelling limits or intrinsic properties of discs is still unknown. The prediction that disc would, thus, show, in their inner hot region, a variety of chemistry, raises the need to determine new sets of dust mixtures that can take into account the various changes in mineralogy, both in terms of global composition and spatial variations, that different disc bulk elemental compositions would bring. 
Having more accurate and reliable dust chemistry models becomes especially relevant with the advent of new instruments that can spatially resolve the very inner disc regions of protoplanetary discs. This is the case of MATISSE, the new spectro-interferometer recently installed at the VLTI \citep{2014Msngr.157....5L}. Combining four of the VLT telescopes (either the 8m Unit Telescopes of the 1.8m Auxiliary Telescopes), this instrument can provide MIR spectroscopic observations, down to the sub-au scale, of the hot and warm dust.\\ 
In this work, we derived synthetic infrared spectra of a fiducial protoplanetary disc model with different dust compositions for the inner disc regions. The dust compositions were computed from different condensation sequences (computed at LTE) associated with a given $C/O$ ratio.
Our aim is twofold:
\begin{itemize} 
\item analysing the effect, on the observables, of different initial inner disc compositions and identifying associated trends and signatures. This in order to determine to which extent the disc initial bulk composition and the associated condensation sequence can be constrained from the observations.
\item providing reliable dust mixtures for the modelling and interpretation of future spatially resolved observations of the inner disc regions.  
\end{itemize}
 The paper is structured as follows: in Section~\ref{discchem} we present the thermodynamic model we use to derive the inner disc solids compositions. In Section~\ref{deriv}, we describe our approach to derive the synthetic spectra and observables associated with the different inner disc compositions. We present the results in Section~\ref{results}, followed, in Section~\ref{observation}, by a first interpretation of inner disc observations using our synthetic spectra. Then, we discuss our results in Section~\ref{discussion}. Finally, in Section~\ref{conclusions}, we summarize our work and provide an outlook for future protoplanetary disc studies.

\section{Disc model and chemistry}
\label{discchem}

To derive our fiducial dust composition for solar and other $C/O$ ratios, we started from the 15 most abundant elements of the solar photosphere as derived by \citet{2009ARA&A..47..481A}. The solar $C/O$ elemental abundances were directly taken from the elemental distribution in \citep{2009ARA&A..47..481A}. The elemental abundances corresponding to the lower and higher $C/O$ ratios were obtained from the solar elemental distribution by changing only the initial amount of C(g) and O(g) to obtain the desired ratio by solving these two simple relations: $C(g) + O(g) = 0.0697$ (see Table~\ref{dustCO} in  \citet{2016MNRAS.457.1359P}) and $C(g)/O(g) = X$  where X is equal to 0.4 or 1.0. These are the values of the $C/O$ ratios for which sensitive changes to the dust chemistry occur \citep{1975GeCoA..39..389L,1999IAUS..191..279L,2012ApJ...757..192J}, and they  cover the uncertainties that modelling the stars' elemental abundances of C and O could bring \citep{2012ApJ...747L..27F}.

From those three different initial elemental abundances, we performed 1D equilibrium calculations using the HSC 8.0 software package\footnote{\href{http://www.hsc-chemistry.net}{http://www.hsc-chemistry.net}} \citep{roine2002outokumpu} in the temperature range of $500< T\rm{(K)}<1850$ and with a constant pressure of $P=10^{-3}$~bar. The lowest temperature value roughly correspond to a distance of 1~au from the protostar \citep{Dalessio1998,1999ApJ...527..893D}. In this work, our focus is on resolving the radial changes in the solid chemistry, implied by different bulk chemistries, rather than retrieving for instance fine chemistry of single objects. A 1D equilibrium calculation model, although the simplest approach, is an appropriate starting approximation for the following reasons: (i) 2D temperature (T(R,Z)) and pressure (P(R,Z)) profiles of typical continuous disc models show a signficantly larger temperature gradient in the radial direction than in the vertical one, at the optically thin disc surface (above the vertical $\tau = 1$ line) where most of the infrared emission and solid-state infrared features originates. The situation changes when considering all the vertical structure of the disc \citep{Dalessio1998,1999ApJ...527..893D, 2016MNRAS.457.1359P}; (ii) dust material located within the optically thick zones, such as the disc midplane, do not contribute to the infrared spectra. Moreover, the presence of a dead zone  \citep[e.g.][]{1996ApJ...457..355G} will suppress the turbulence and thus likely impair mixing processes between the disc upper layers and deep layers; (iii) Furthermore, for our study cases, the chemistry  of the dust would not be  affected by removing a dimension because, for each considered C/O ratio, the dust  is largely dominated by few key-species (see table 1 in the manuscript). These species are stable in a wide temperature and pressure ranges and there are basically no variations in the dust composition at different low pressures  \citep[e.g.][]{1995GeCoA..59.3413Y,2011MNRAS.414.2386P}. 

To summarize, 1) the larger radial temperature gradient in the optically thin disc regions, 2) the large dominance of key species in term of mass fraction, and 3) the independence of the main dust components on pressure in the considered disc regions, make the use of 1D-fixed pressure a reliable starting point to study the effects of different C/O ratios in infrared spectra.

Nevertheless, the effect  of pressure could become important when moving very close to the innermost region, where silicate and iron will start to vaporise leaving the most refractory component as the only solids. Note that detecting this effect would require spatially resolving the very innermost region of the disk so to exclude the silicates and iron from the probed dust mixture.

The equilibrium calculations were performed in bins separated by steps of 10 K in the selected temperature range. Two temperature zones, encompassing the $R<1$~au region of the disc, were considered : a high-temperature zone (($T>1000$~K) and a low-temperature zone ($500<T\rm{(K)}<1000$). T$\sim$1000~K is the average temperature in which annealing and total equilibrium can occur in short timescales and where sharp transition in the dust chemistry in discs occurs \citep{1995GeCoA..59.3413Y,2004A&A...413..571G,2011A&A...529A.111R,2016MNRAS.457.1359P}. We then computed the average dust composition within these two zones., i.e. we summed all the abundances of the solids present in each bin of the selected zone and derive the averaged  wt\% for each single component. These  averages constitute our dust mixtures for the high and low temperature disc regions respectively.

 Dust in the outer cooler disc region ( $T<500$~K, $R>1$~au) will be mainly characterised by aggregates that never experienced high temperature. Our fiducial dust composition for the outer disc will, thus, be 90\% amorphous astrosilicates + 10\% graphite following \cite{2011ApJ...734...51O}. This composition is the same for all the $C/O$ ratios. Although a crude approximation, keeping the composition of the outer disc constant will help us to disentangle the effect of the different chemistries in the warmer inner disc region. Moreover, when resolving the inner disc regions, the contribution of the cold outer disc will tend to fade out in the measurements provided by long-baseline interferometry (see Section~\ref{deriv}).

In Table~\ref{dustCO} we report the resulting normalized composition (wt\%) of the dust mixture in each temperature region and for each considered $C/O$ ratio. Only are shown the major components for which optical data are available in the literature. These are the considered species for the radiative transfer modelling (see section 3), which requires such optical data to compute the dust opacity. All the other components have minor abundances (generally on average less than $\sim5\%$). The only major component for which we could not find optical data is iron silicide (FeSi with a molecular weight of 83.93 g/mol). In our dust opacity computation, we replaced it by pure iron knowing that iron amounts to 66.54 \% of the total mass. As expected, different $C/O$ ratios are characterised by a specific dust chemistry. For the supersolar $C/O$ ratio we see the disappearance of silicates in the high temperature region, and the appearance of graphite and large amount of FeSi. For subsolar $C/O$ ratios, the dust chemistry is mostly oxide (\ce{FeO}, \ce{MgO} and \ce{SiO2}) and hydro-oxide dominated. This is in agreement with previous theoretical calculations \citep{1975GeCoA..39..389L,1999IAUS..191..279L,2012ApJ...757..192J}.  


\begin{table*}													
\centering													
\caption{Fiducial dust distribution (in wt\%) for the high ($T>1000$~K) and low ($500<T<1000$) temperature regions, at $C/O=0.4$, solar (0.54) and 1. Major compounds for which optical data are available are shown. The last column indicate the original reference for the optical data.}													
\begin{tabular} {l c c c c c c c c c}													
\hline													
$C/O$	&	0.4	&	0.4	&	Solar	&	Solar	&	1	&	1 &	\\
\hline													
T	&	high	&	low	&	high	&	low	&	high	&	low	& optical constant (ref)\\
\hline													
forsterite (\ce{Mg2SiO4})	&	13.06	&	21.14	&	25.57	&	39.13	&	4.12	&	50.21 &	\cite{2006MNRAS.370.1599S}\\
enstatite (\ce{MgSiO3})	&	23.74	&	15.97	&	38.32	&	20	&	x	&	14.33 &	\cite{Jaeger1998} \\
metal (\ce{Fe})	&	1.35	&	0.86	&	36.11	&	27.71	&	7.3	&	29.9 & \cite{Ordal85}	\\
troilite (\ce{FeS})	&	x	&	x	&	0	&	10.72	&	x	&	x &	\cite{Henning1996} \\
fayalite (\ce{Fe2SiO4})	&	x	&	x	&	$\sim$0	&	2.44	&	x	&	4.85 &	\cite{Fabian2001a}\\
\ce{FeSi}	&	x	&	x	&	x	&	x	&	87.28	&	x &	\cite{Ordal85}\\
													
graphite (\ce{C})	&	x	&	x	&	x	&	x	&	1.23	&	0.71 &	\cite{1993ApJ...402..441L}\\
\ce{FeO}	&	43.8	&	53.91	&	x	&	x	&	x	&	x &	\cite{Henning1995}\\
\ce{MgO}	&	8.56	&	5.59	&	x	&	x	&	x	&	x &	\cite{Henning1995}\\
\ce{SiO2}	&	9.25	&	1.58	&	x	&	x	&	x	&	x &	\cite{Zeidler2013}\\
spinel \ce{MgAl2O4}	&	0.24	&	0.94	&	x	&	x	&	x	&	x &	\cite{Fabian2001b}\\
\hline													
\end{tabular}													
\label{dustCO}													
\end{table*}													

\section{Deriving the IR spectra and interferometric observables}
\label{deriv}
\begin{figure}
 \resizebox{\hsize}{!}{\includegraphics{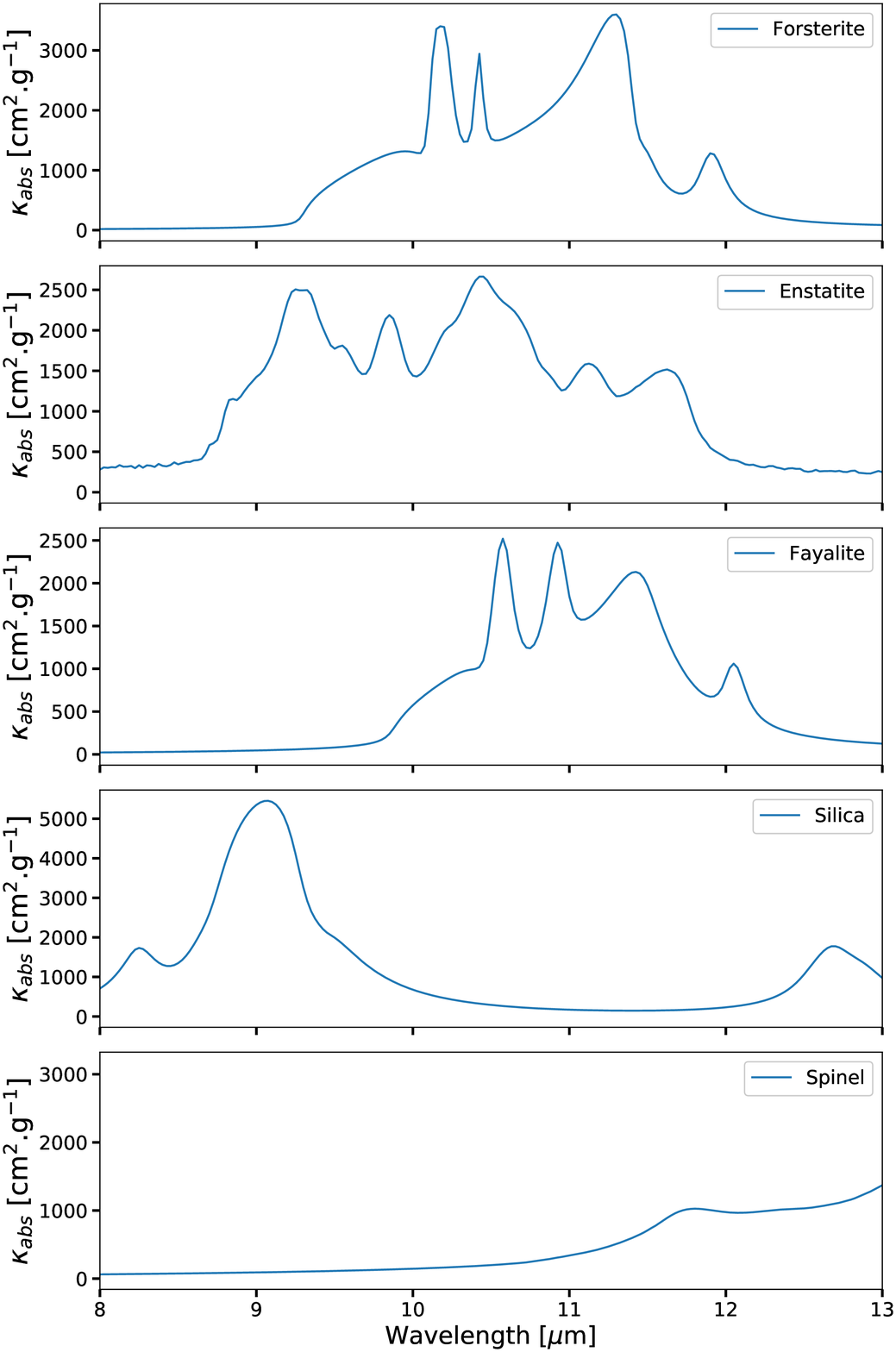}}
  \caption{ Mass absorption coefficients of the dust species presented in Table~\ref{dustCO}, which present features in the 8 to 13 $\mu$m spectral domain. Those coefficients were computed using a DHS approach (see Section 3).}
 \label{fig:opac}
\end{figure}

From our fiducial dust compositions, the computation of the synthetic IR spectra and interferometric observables was performed using radiative transfer. We defined a disc model setup in several steps : 

\paragraph*{Dust density distribution}
We considered the same 2D axisymmetric model taken from \citet{Dalessio1998,1999ApJ...527..893D}. This flared disc model provides the radial and vertical dust density distribution, denoted as $\rho(r,\theta)$.
\paragraph*{Dust opacity}
The opacity (absorption and scattering properties) associated with the dust compositions derived in Section~\ref{discchem}, is computed from the optical constants of the corresponding dust species. Indeed, dust grain radiative properties are described by their optical constants mostly taken from the Jena database\footnote{Available at \href{http://www.astro.uni-jena.de/Laboratory/OCDB/}{http://www.astro.uni-jena.de/Laboratory/OCDB/}}. The original references are indicated in Table~\ref{dustCO}. In a dust aggregate, each monomer is composed of roughly one material component, which is then mixed with other monomers of the same composition or not, to make up the aggregate. Mixing between different material components was computed using the effective medium theory \citep[see e.g.,][and references therein]{2016A&A...585A..13M}. In such effective medium computations, the refractive index of the composite particle (aggregate) is computed from the refractive indices of the submaterials weighted by their abundances. The mass absorption and scattering coefficients (in cm$^{2}$.g$^{-1}$), $\kappa_{\rm abs}(\lambda),$ and $\kappa_{\rm sca}(\lambda)$ were then computed assuming the grains are irregular. Such an irregularity is approximated by a Distribution of Hollow spheres \citep[DHS, ][]{2005A&A...432..909M} with a maximum vacuum fraction of f$_{\rm max}=0.8$. In the DHS approximation, f$_{\rm max}$ is the `irregularity' parameter since there is no observable difference between porosity and irregularity. Finally, assuming a grain size distribution $n(a)\propto a^{-3.5}$ \citep{1977ApJ...217..425M}, with $a_{\rm min}$ and $a_{\rm max}$ the minimum and maximum grain sizes, 
the size-averaged mass absorption/scattering coefficient was obtained by adding the mass absorption/scattering coefficients of each grain size times their mass fraction. Figure \ref{fig:opac} illustrates the computed mass absorption coefficients of the dust species used in our work and that present spectral features in N-band. 
For each radial zone of the disc, i.e. $r<1$~AU and $r>1$~AU, the composition is assumed homogeneous in the vertical direction, as already mentioned in Section~\ref{discchem}. 
We consider a single maximum grain size ($a_{max}=1~$mm), and two minimum grain sizes ($a_{min}=0.05~\mu$m and $a_{min}=1.0~\mu$m) to explore the effect of such parameter on the radial changes impacting the inner disc N-band spectra. 

\paragraph*{Cristallinity}
The cristallinity degree of dust grains in the innermost disc regions ($\lesssim 10$~au) could only be addressed so far by mid-infrared spectro-interferometry of the N-band silicate solid-state features. For instance, \citet{2004Natur.432..479V} found inner disc cristallinity fractions between 40\% and 100\%. In our work, we assumed 100\% crystallinity for the inner disc dust to highlight the differences brought by the different $C/O$ chemistries and build up a case study. Such an assumption is physically motivated by the fact that in the innermost regions, temperatures are high, thus allowing direct condensation from gas and fast annealing and crystallization. Combined with turbulence in the active disc surface layers, which mixed the grains radially and vertically, all grains could reach in theory the equilibrium state driven by high temperatures. Finally, the involved evolution timescales (we are considering objects which are about 1 million years old) should allow a complete remixing and crystallization within those regions. 

\paragraph*{Temperature distribution, synthetic spectra and images}
From the dust density distribution and a given opacity, we recomputed the temperature distribution of the disc model using the Monte-Carlo radiative transfer code RADMC3D\footnote{Publicly available at \href{http://www.ita.uni-heidelberg.de/dullemond/software/radmc-3d/}{http://www.ita.uni-heidelberg.de/dullemond/software/radmc-3d/}}. From the computed radiation field and temperature distribution, RADMC3D can also produce synthetic spectra and images by integrating the radiative transfer equation along rays (ray-tracing method). Isotropic scattering was considered in the modeling.

For each dust composition, we produced synthetic spectra considering specific spectral coverages and resolutions associated with two instruments : the {\itshape Spitzer Space Telescope}, especially its Infrared Spectrograph IRS \citep{2004ApJS..154...18H}, and the second generation VLTI instrument MATISSE \citep{2014Msngr.157....5L}. This choice was driven by the fact that these two instruments can serve as an observational basis for our work. Indeed, Spitzer produced an extensive database of (spatially unresolved) MIR spectra of protoplanetary discs, while the new MIR spectro-interferometer MATISSE can perform spatially resolved (down to the sub-au scale) spectroscopic observations of warm dust in the $\sim 0.1-10$~au region of discs. 
 
For each dust composition ($C/O=0.4$, $C/O=0.54$, $C/O=1$), we produced spatially unresolved synthetic spectra:
\begin{itemize}
\item a broadband optical and IR SED, ranging from 0.1 to 35~$\mu$m in wavelength. 
\item Three synthetic Spitzer spectra : a low resolution ($R\sim100$) spectrum from 5.2 to 14.5~$\mu$m, and two high resolution ($R\sim600$) spectra, from 9.9 to 19.5~$\mu$m, and from 18.7 to 37.2~$\mu$m.
\end{itemize}

As a parallel model outcome, we also produced synthetic N-band spatially resolved (or correlated) spectra according to the characteristics of MATISSE (Field-of-view of about 600~mas when using the 8m Unit Telescopes, medium spectral resolution of R=220, spectral coverage from 8~$\mu$m to 13~$\mu$m). The correlated flux measured by an interferometer is the spatially and temporally coherent flux collected by the different apertures, which will interfere to form the fringes. It includes the flux contributions from the spatially unresolved regions of the astrophysical source. 
As the emission of circumstellar discs is strongly centrally peaked \citep{2015A&A...581A.107M}, the correlated spectra measured by a MIR interferometer are dominated by the inner few au of the disc. At first order, as the interferometer baseline length (i.e. the distance between the telescopes) increases, the measured correlated flux will thus originate from disc regions closer and closer to the central star. Determining the disc regions that are actually contributing to the correlated flux measured for a given baseline length usually requires to consider a specific geometry. Indeed, for a given couple of telescopes, according to the Van Cittert-Zernike theorem, the correlated flux corresponds mathematically to the modulus of the Fourier transform of the brightness distribution of the object measured at the angular spatial frequency $B/\lambda$, where B is the distance between the telescopes.  For an interferometer observing a narrow ring-like brightness distribution, as in the case of the inner disc regions, the angular resolution criterion can be taken as $\sim \frac{0.77\lambda}{B}$, where $B$ is baseline length projected onto the plane of the sky\footnote{The first zero of the Fourier transform of a infinitesimally thin ring of angular diameter $\theta$ occurs at an angular frequency equal to $0.77/\theta$.}. This roughly means that the disc regions located inside a central region of about $\frac{0.77\lambda}{B}$ in angular diameter will contribute to the measured correlated flux. According to this criterion, our synthetic correlated spectra thus `probes' disc regions closer than about $\frac{0.77\lambda}{2B}$ to the central star. 
Here, we computed synthetic correlated N-band spectra for three different baseline lengths (B=50m, B=100m, and B=130m), where B=130m represents the maximum baseline length achievable at the VLTI with the 8m UTs. In our radiative transfer model, the disc is considered to be at a typical distance of 100 pc. At such distance, the corresponding disc regions that are probed are $r < 1.5$~au, $r < 0.8$~au, and $r < 0.6$~au, respectively. In the following, we refer to the correlated spectra as the `inner-disc' spectra. 
Depending on the radial variation of the disc mineralogy, changes can thus be expected in the N-band solid-state features in the inner-disc spectra. To assess the detectability of such changes, we computed, from our synthetic spectra, normalized continuum-subtracted spectra following \citet{2005A&A...437..189V}. First, the N-band continuum $F_{\rm cont} \left(\lambda \right)$ is estimated, for each spectrum, through a linear fit from 8.5~$\mu$m to 12.3~$\mu$m, which is the spectral width of the solid-state emission band. Then the normalized continuum-subtracted spectrum is computed as follows: 
\begin{equation}
F_{\rm norm} \left(\lambda \right)=1+\frac{F \left(\lambda \right)-F_{\rm cont} \left(\lambda \right)}{<F_{\rm cont}>},
\end{equation}
where $<F_{\rm cont}>$ is the mean of the N-band continuum. With this definition, $F_{\rm norm}$ preserves the shape of the emission band while being independent from the underlying N-band continuum spectral slope and level. In this way, the different spectral features in the inner-disc spectra can be directly compared. Finally, error bars associated with the synthetic MATISSE data were generated using an IDL simulation tool, which is based on a noise model detailed in \citet{2016SPIE.9907E..28M}. Here, we considered a source brightness of about 5 Jy in N-band, which is typical for a significant number of Herbig stars and for rather bright T Tauri stars.  

\section{Results}
\label{results}

\subsection{spatially unresolved spectra}

\begin{figure*}
 \centering
 \resizebox{\hsize}{!}{\includegraphics[width=80mm,height=50mm]{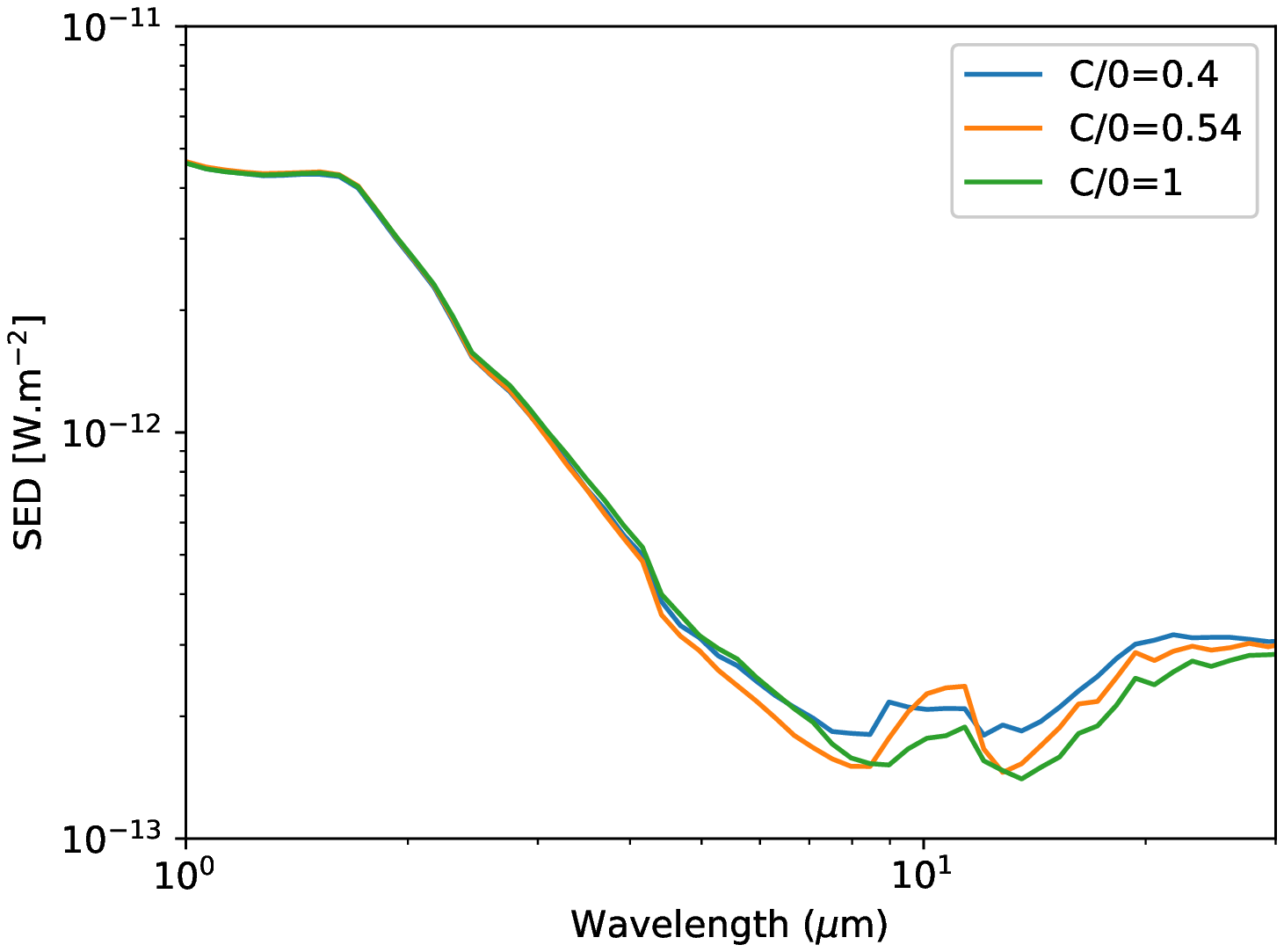}
 \includegraphics[width=80mm,height=50mm]{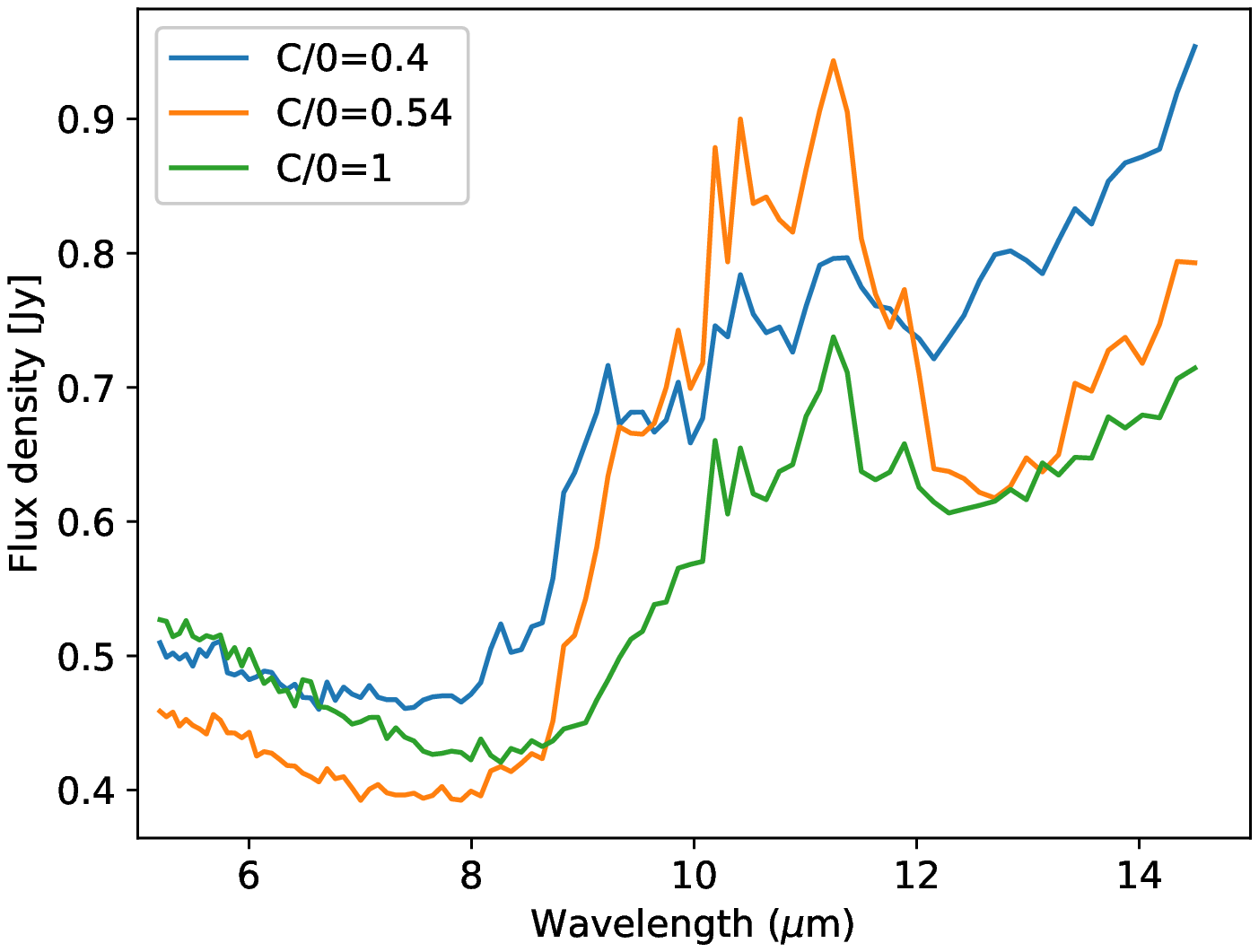}}\\
 \resizebox{\hsize}{!}{\includegraphics[width=90mm,height=55mm]{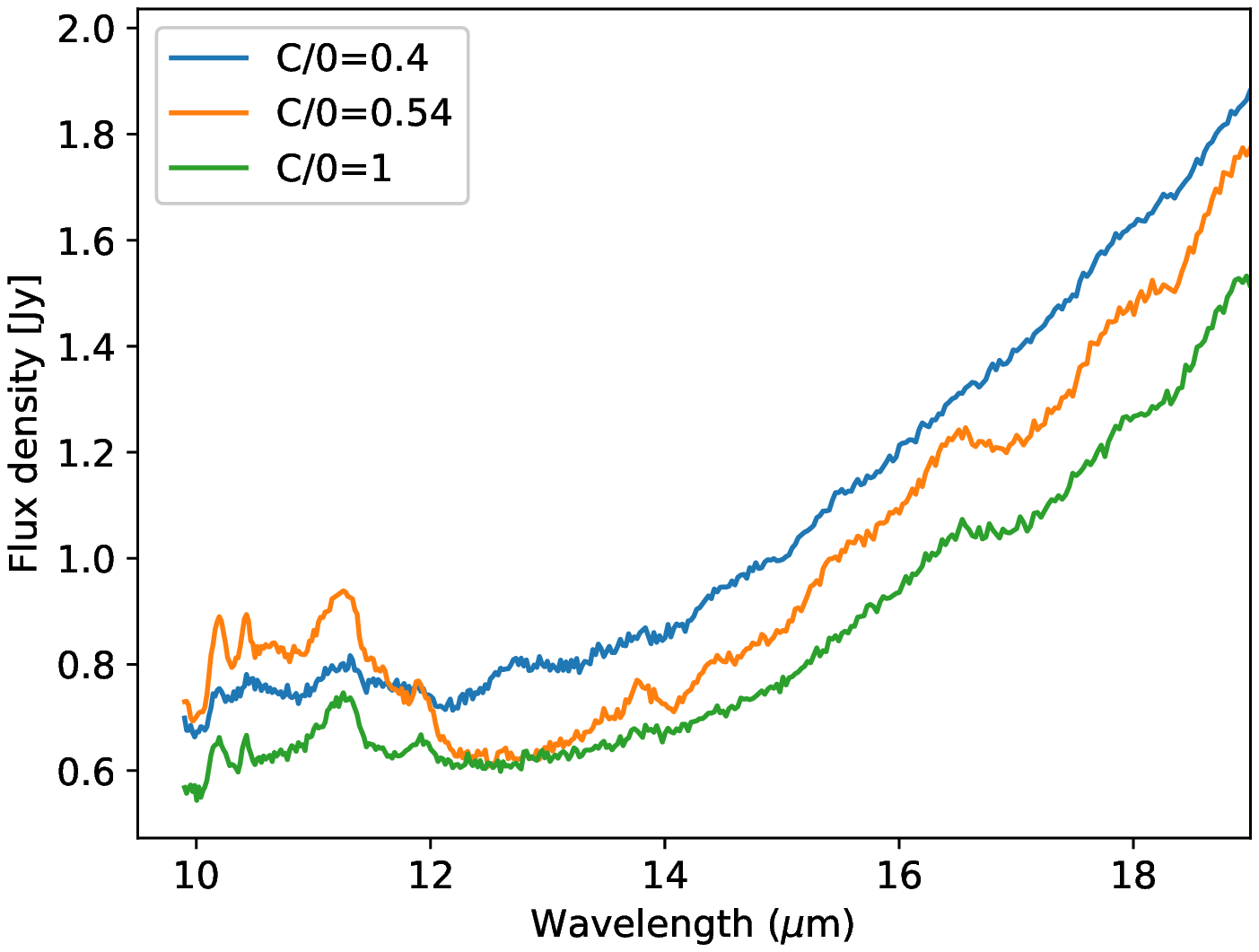}
 \includegraphics[width=90mm,height=55mm]{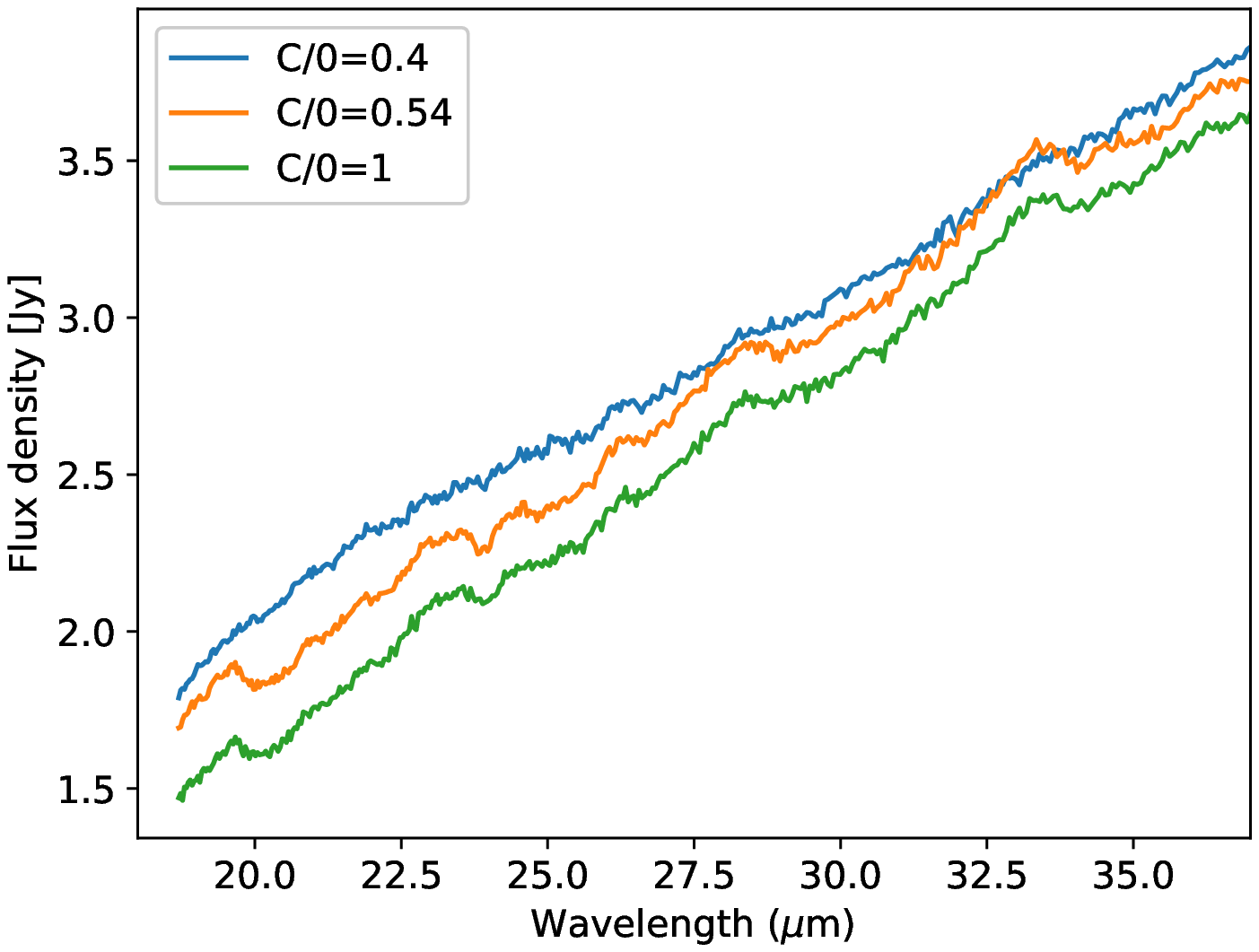}}
  \caption{ {\itshape Top left}: Synthetic broadband optical and IR SED {\itshape Top right}: Synthetic low resolution ($R\sim100$) SPITZER spectrum from 5.2 to 14.5~$\mu$m. {\itshape Bottom left}: synthetic high resolution ($R\sim600$) spectra, from 9.9 to 19.5~$\mu$m. {\itshape Bottom right}: synthetic high resolution ($R\sim600$) spectra from 18.7 to 37.2~$\mu$m. 
  }
 \label{fig:radmcmodel_unres}
\end{figure*}

We present in Fig.\ref{fig:radmcmodel_unres} four synthetic spatially unresolved spectra. They consist of one broadband IR SED, and three SPITZER spectra computed at two different spectral resolutions ($R\sim100$, and $R\sim600$).

In the solar $C/O$ case, all the synthetic unresolved spectra show a broad silicate band partly due to the enstatite fraction in the inner disc ($<1$~au) and the amorphous silicates in the outer disc ($>1$~au). As seen in the lower left and top right panels of Fig.\ref{fig:radmcmodel_unres}, the crystalline silicate peaks due to the forsterite- and enstatite-rich regions clearly appear on top of the amorphous band. On one hand, as the $C/O$ ratio increases ($C/O=1$), the global trend is a decrease of the amplitude of the broad silicate band and the crystalline features (see lower left and top right panels of Fig.\ref{fig:radmcmodel_unres}). We can also note in the broadband SED (top left panel of Fig.\ref{fig:radmcmodel_unres}) a slight increase of the NIR excess due to the larger fraction of elements having large NIR opacity such as Iron and Graphite. On the other hand, a lower $C/O$ value ($C/O=0.4$) implies a decrease and flattening of the silicate emission band (see top right panel of Fig.\ref{fig:radmcmodel_unres}). As expected, the forsterite feature at 11.3~$\mu$m is reduced in favor of the short edge ($\sim 9$~$\mu$m) of the silicate band, which represents the contribution from enstatite and silica. At longer wavelengths $> 20 \mu$m (see bottom right panel of Fig.\ref{fig:radmcmodel_unres}), the $C/0=0.4$ case produces basically a featureless spectrum with residual enstatite features while the two other spectra still show the enstatite and mainly the forsterite features.
The change in $C/O$ ratio has thus a non-negligible impact on the infrared part of the spatially unresolved spectra.

\subsection{Spatially resolved spectra}
\begin{figure*}
 \centering
 \includegraphics[width=80mm,height=50mm]{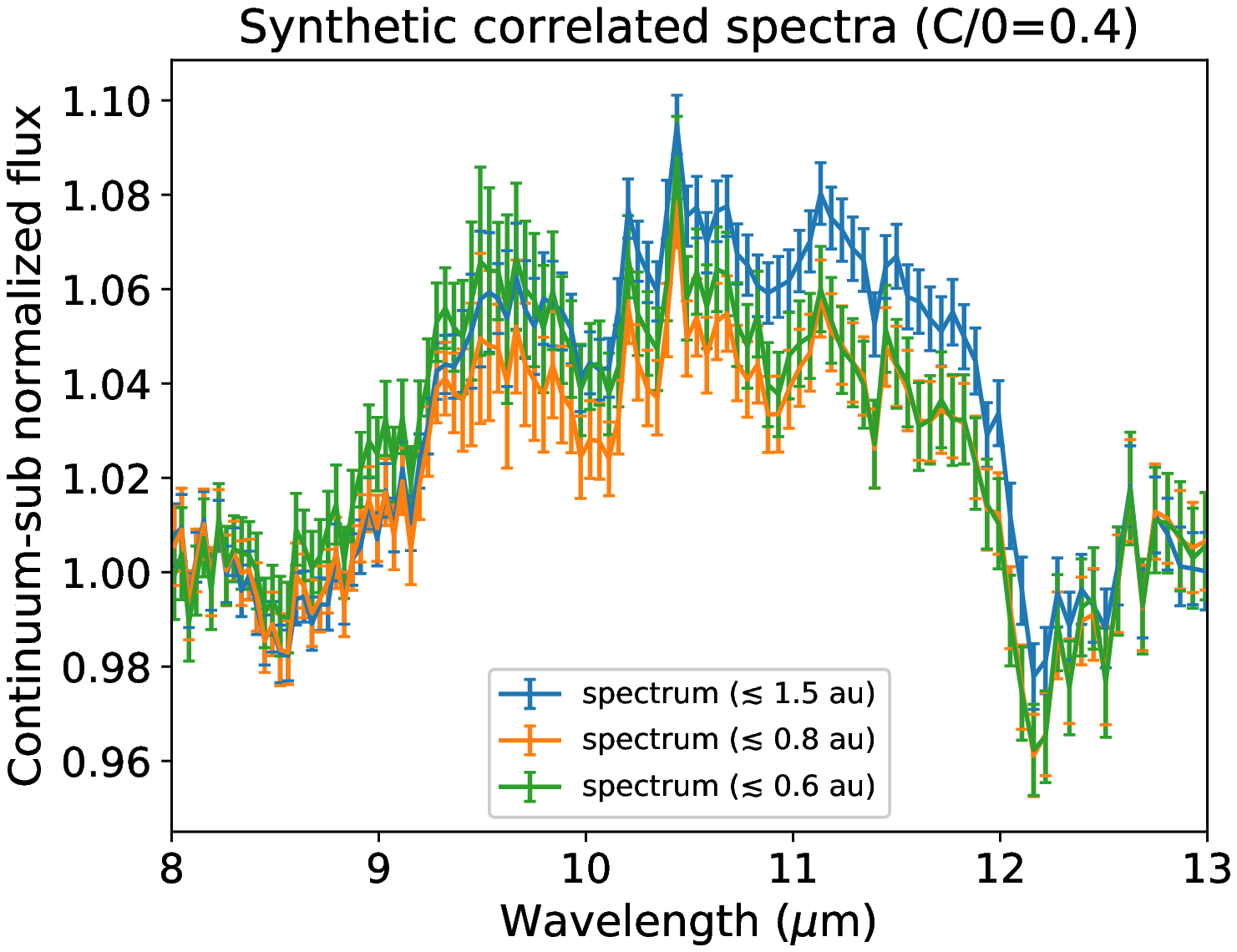}
   \includegraphics[width=80mm,height=50mm]{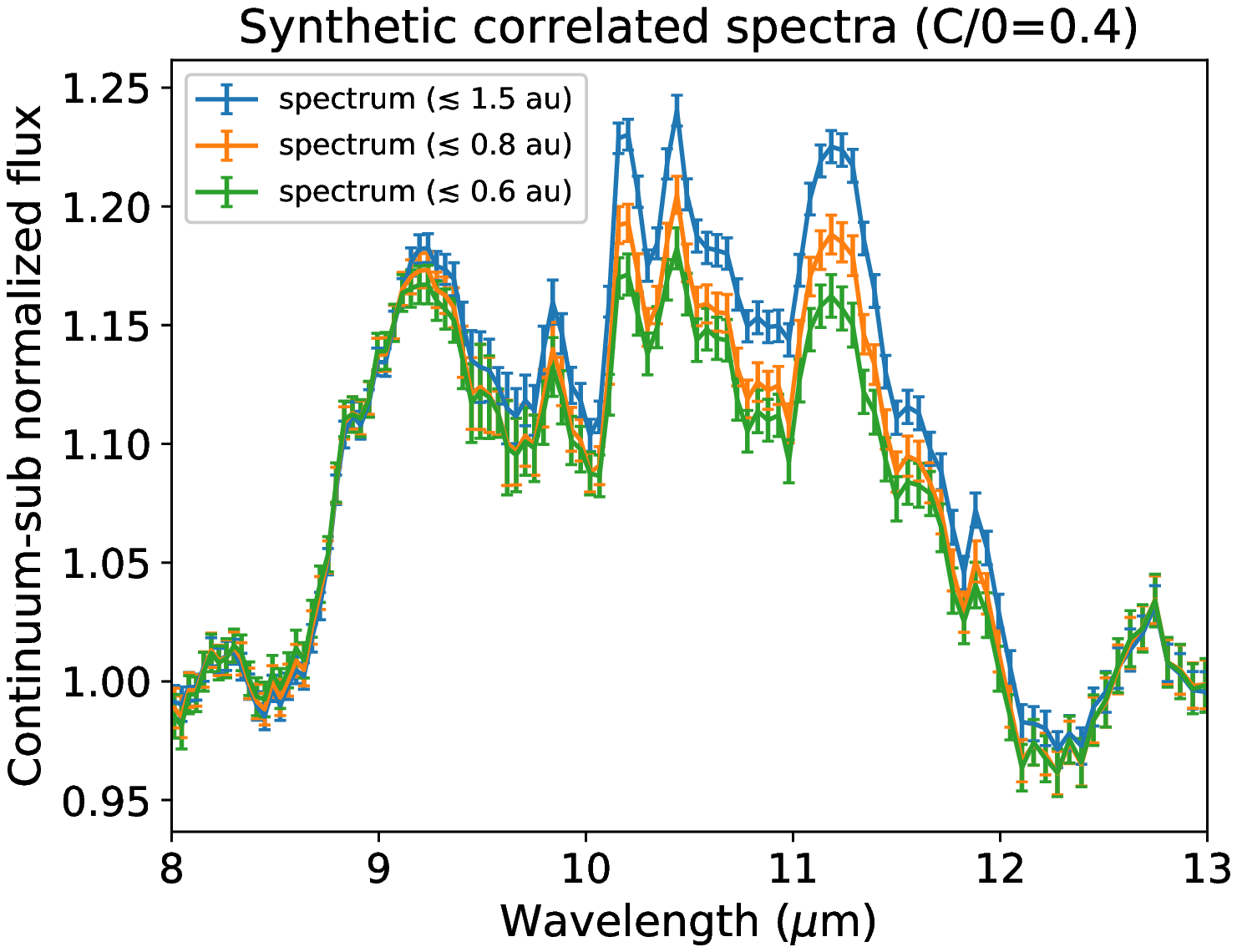} \\
 \includegraphics[width=80mm,height=50mm]{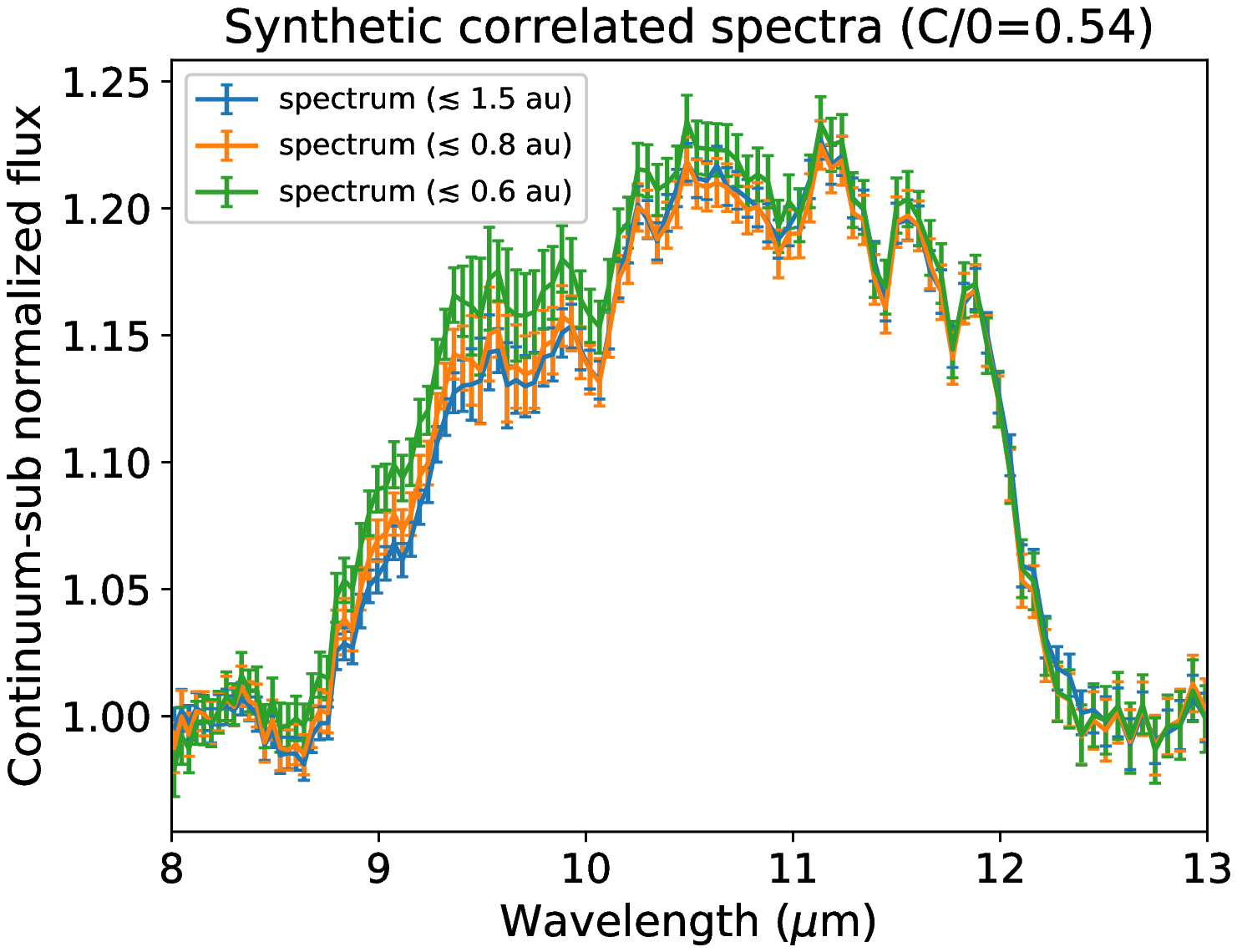}
  \includegraphics[width=80mm,height=50mm]{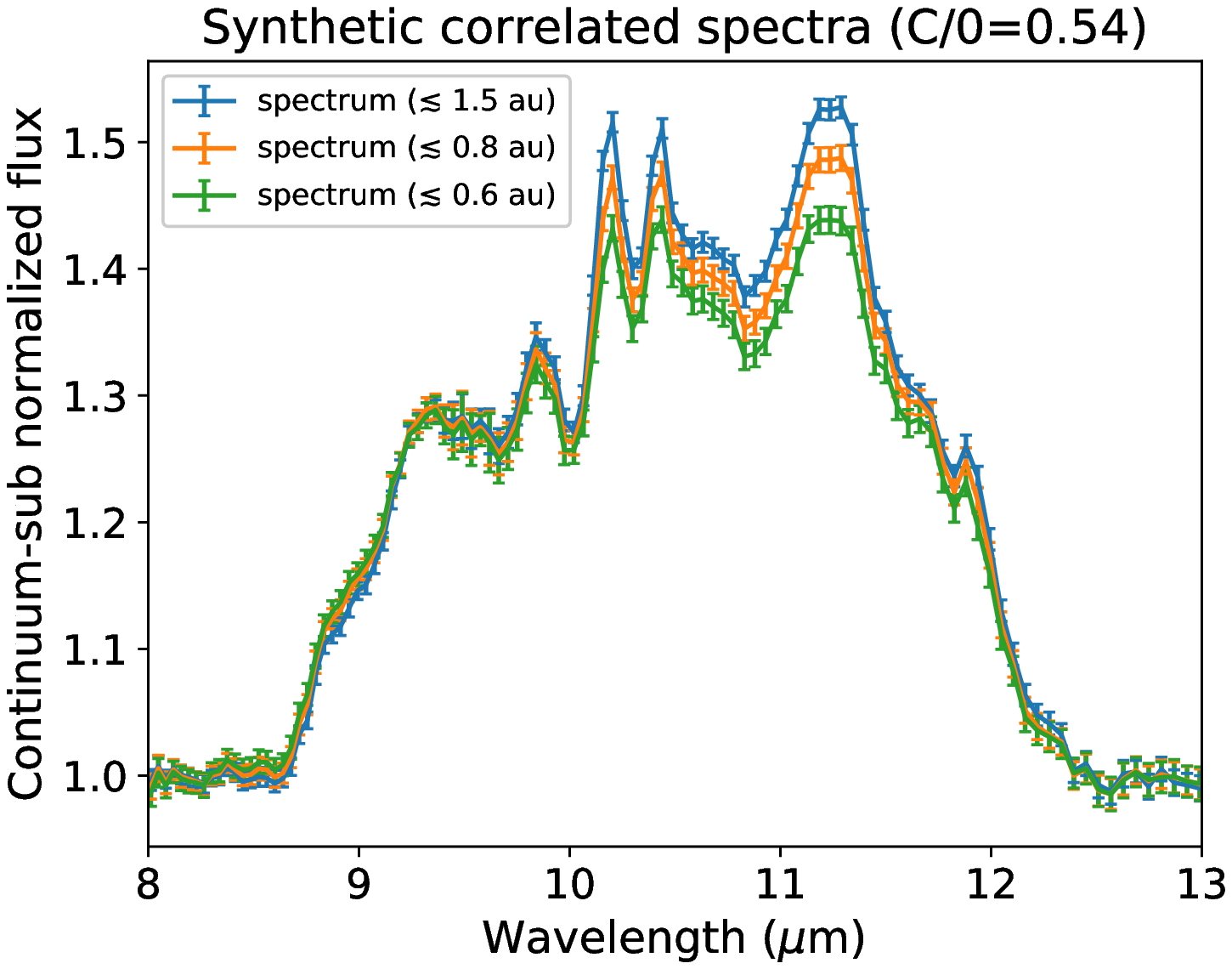}\\
 \includegraphics[width=80mm,height=50mm]{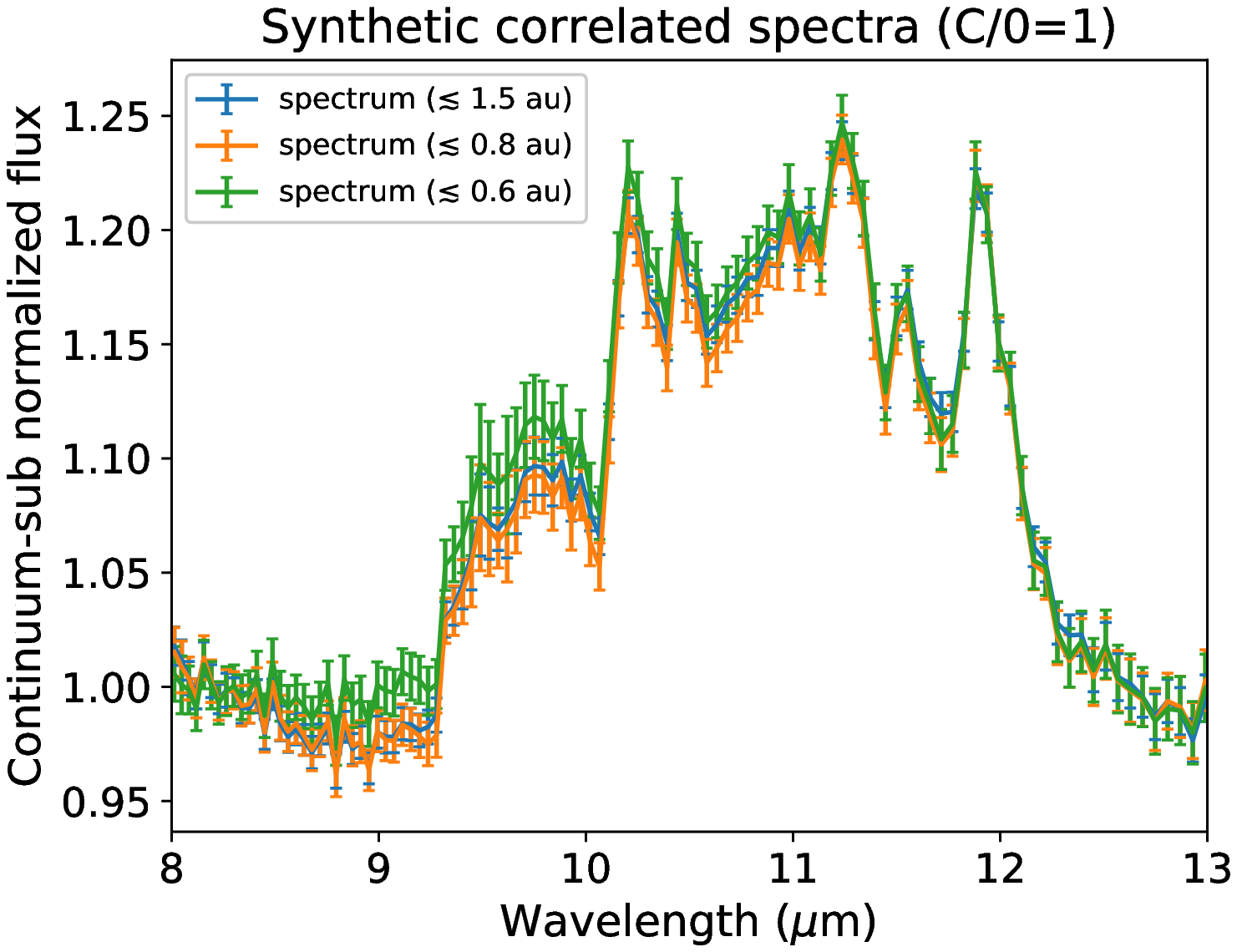}
 \includegraphics[width=80mm,height=50mm]{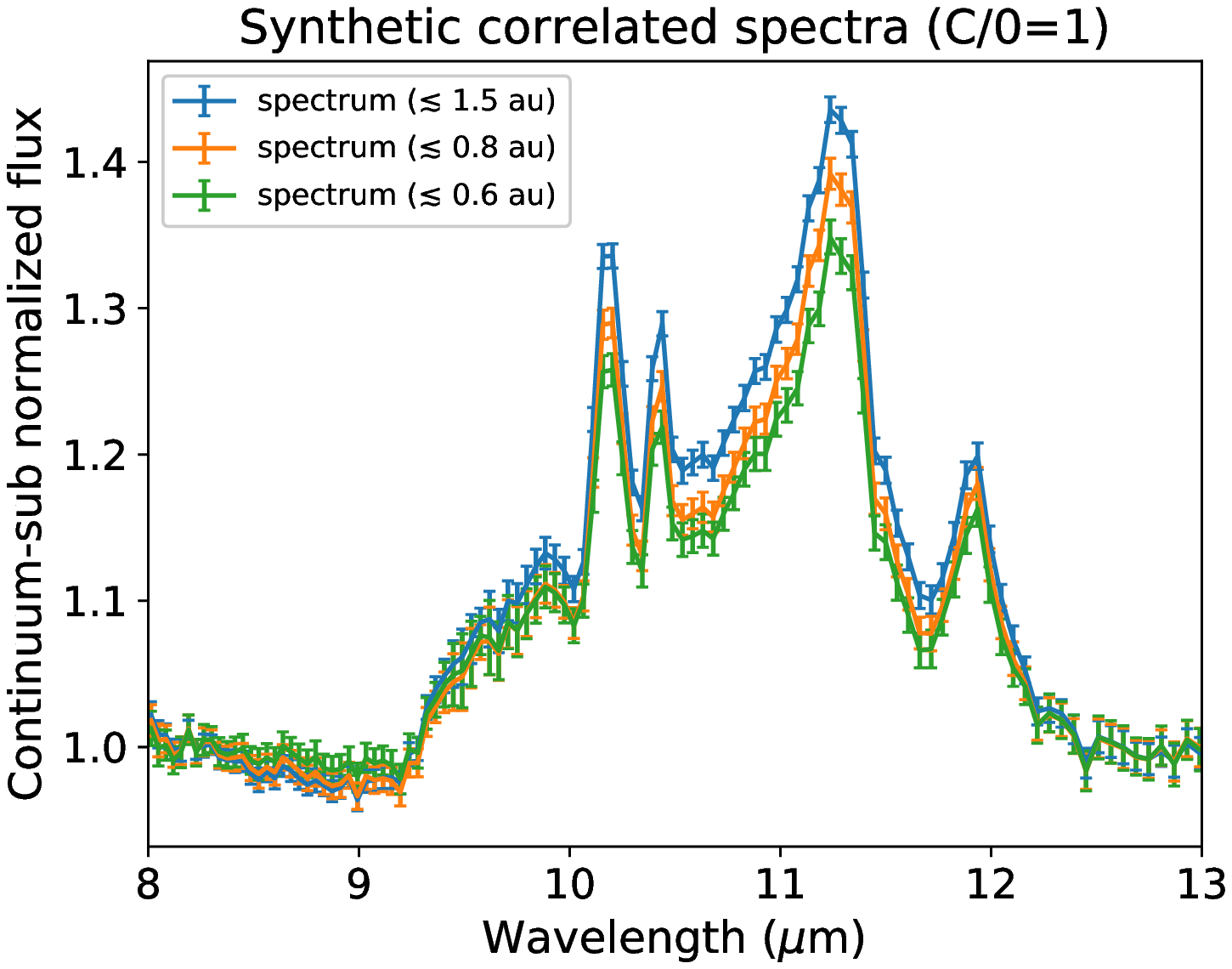}\\
  \caption{From Top to Bottom: synthetic correlated MATISSE spectra at high spectral resolution (R=220) for a $C/O=0.4$, $C/O=0.54$, and $C/O=1$ composition, respectively. The left column plots correspond to $a_{min}=0.05~\mu$m, and the right column to $a_{min}=1.0~\mu$m. The three correlated spectra probes the inner 2 au, 1 au, and 0.6 au of the disc (from the central star), respectively.}
\label{fig:radmcmodel_res}
\end{figure*}

We present in Fig.\ref{fig:radmcmodel_res}, for every dust composition, the synthetic correlated N-band spectra.
As mentioned before, the Solar $C/O$ case produces a low-T Forsterite-rich region and a high-T Enstatite-rich region. The synthetic MATISSE spectra for $a_{min}=0.05~\mu$m show such an expected decrease of the amplitude of the Forsterite peaks (two peaks between 10 and 10.5~$\mu$m and one peak at 11.3~$\mu$m) as we move inward. Relatively, the amplitude of the Enstatite features (between 9 and 10~$\mu$m, and at 10.5~$\mu$m) is maintained. On the other hand, for  $a_{min}=1.0~\mu$m,  the global amplitude of the emission band is lower, as expected, and the relative radial change in the chemistry is evidenced by the slight increase of the enstatite features. There is no clear decrease of the forsterite peaks as we move inward. 

In the $C/O=0.4$ case, calculations return silica as stable phase and less forsterite than in the solar $C/O$ case. This is evident in the spectra, especially for $a_{\rm min}=0.05~\mu$m. As we probe the inner 1 au, we see a significant decrease of the forsterite peaks. The contribution of silica appears at $\sim 9 \mu$m, together with a steady contribution from enstatite. For the innermost spectrum (in green), the 9~$\mu$m silica-enstatite part becomes more prominent than the long-wavelength edge ($\lambda > 10$~$\mu$m). That makes the $C/O=0.4$ spectra clearly different from the solar case, where the long-wavelength edge ($\lambda > 10$~$\mu$m) is always higher. 
Moreover, the presence of a significant fraction ($\sim 50$\%) of iron oxide, which is featureless in the mid-infrared, in the inner 1~au decreases by about 15\% the global amplitude of the emission band, compared to the solar case.  

In the $C/O=1$ case, the lack of silicates, as we move inward in the inner 1~au region, induces a decrease of the forsterite peaks. This is especially visible for $a_{min}=0.05~\mu$m. However, it is not correlated with an increase of the enstatite features. Indeed, the emission band between 9 and 10~$\mu$m is very weak and does not change with the distance to the star. It is expected since the high-T region in the $C/O=1$ case contains essentially ferrosilicon (90\%) and a bit of graphite (1\%), which are essentially featureless species in the IR domain covered by MATISSE. In the $a_{min}=1~\mu$m case, the crystalline features are damped and no radial variation can be clearly detected in the spectrum.  

In conclusion, a different global bulk chemistry can imprint clear changes in the amplitude and shape of the N-band emission band for both minimum grain size values. Those changes appear even more visible as we resolve the innermost 1 au region.  

\section{Observational comparison}
\label{observation}
\subsection{The sample}
\begin{table*}													
\centering													
\caption{Stellar parameters along with the parameters of the interferometric observations used for this work. The distances were taken from the DR2 Gaia catalogue \citep{2018yCat.1345....0G}. The spectral types and stellar luminosities were taken from \citet{2018A&A...617A..83V} and references therein.}													
\begin{tabular} {l c c c c c c c c }													
\hline													
Name	&	Type	& Spectral	&	d	&	$L_{\star}$	&	Time	&	Base &$B_{\rm p}$& $PA$	\\
	&		&	type	&	[pc]	&	$L_{\sun}$	&	UTC	& &[m] & [$\circ$]	\\
\hline													
HD 142527	&	HAe	&	F6	&	$157\pm1$	&	$21.5\pm3.6$ &2003-06-14T00:32	& UT3-UT1	& 102.2 & 10.9	\\
	&	&	&	&   &2003-06-14T00:54	& UT3-UT1	& 102.1 & 14.3	\\
		&	&	&	&   &2012-06-06T00:16	& UT2-UT4	& 77.3 & 41.9	\\
\hline		
HD 144432	&	HAe	&	F0	&	$155\pm1$	&	$43\pm10$ &2003-06-17T04:46	& UT2-UT4	& 98.6.2 & 41.1	\\
	&	&	&	&   &2003-06-17T04:51	& UT2-UT4	& 98.2 & 41.5	\\
		&	&	&	&   &2003-06-17T05:07	& UT2-UT4	& 97.0 & 42.5	\\
		&	&	&	&   &2004-04-11T05:55	& UT3-UT2	& 46.0 & 28.1	\\
\hline		
HD 163296	&	HAe	&	A1	&	$101.5\pm1.2$	&	$34.3\pm7.1$ &2003-06-14T03:13	& UT3-UT1	& 99.4 & 17.7	\\
	&	&	&	&   &2010-05-21T08:52	& I1-E0 (AT)	& 60.6 & 122.2	\\
		&	&	&	&   &2010-05-21T09:03	& I1-E0 (AT)	& 59.3 & 124.1	\\
\hline		
AS 209	&	TT	&	K4	&	$121\pm1$	&	$2.23\pm0.34$ &2014-04-16T09:27	& UT3-UT4	& 56.6 & 118.9	\\
	&	&	&	&   &2014-04-16T09:56	& UT3-UT4	& 53.8 & 122.6	\\
	&	&	&	&   &2014-06-18T04:17	& UT1-UT3	& 102.1 & 35.5	\\
	&	&	&	&   &2014-06-18T04:26	& UT1-UT3	& 102.3 & 36.2	\\

\hline													
\end{tabular}													
\label{star_obs}													
\end{table*}

In this section, our aim is to make a first qualitative comparison between our synthetic spectra and actual spatially resolved spectra of the first few astronomical units of discs. We want to assess the possibility of identifying compositional trends in terms of $C/O$ ratio on a few selected sources.
So far, the only instrument that could provide spatially resolved N-band spectra of the inner region of discs is MIDI (Leinert et al., 2003), the former mid-infrared VLTI instrument, which was decommissioned in 2014. MIDI could observe a significant set of both intermediate (Herbig AeBe) and low-mass (T Tauri) young stars \citep[see e.g.,][]{2015A&A...581A.107M,2018A&A...617A..83V}, mostly in low spectral resolution (R$\sim$30). One of the few studies focussing on the au-scale radial variations of the silicate emission band was performed by van Boekel et al. (2004). We thus followed their work and re-analysed qualitatively the set of spectra they presented. This set of spectra corresponds to three Herbig Ae stars: HD142527, HD144432, and HD163296 (see Table~\ref{star_obs}). The calibrated MIDI data for those sources were taken directly from the Optical Interferometry Database of the Jean-Marie Mariotti Center \footnote{Available at \href{http://oidb.jmmc.fr}{http://oidb.jmmc.fr}}.

To complement our new analysis, we also added a lower-mass T Tauri star, AS~209, for which we expect changes in the disc structure and dust properties relative to hotter and more massive Herbig stars. We obtained our AS 209 MIDI data in service mode on 16 April 2014 with the UT3-UT4 baseline and on 18 June 2014 with the UT2-UT3 baseline. For each observing epoch, we obtained two fringe and photometry measurements using the HIGH\_SENS mode of MIDI. In this mode, the photometry (or total flux) is measured just after the fringe acquisition. Fringes and photometry were dispersed with a spectral resolution of $R=30$. In this work, we do not make use of the total MIDI spectrum and consider only the correlated (or inner disc) spectra. We averaged the two measurements for each observing epoch. Our observing program also included a spectrophotometric and interferometric calibrator chosen using the Calvin tool provided by ESO. The calibrator provided an absolute calibration of the correlated flux spectra of AS~209. 

\begin{figure*}
 \centering
 \includegraphics[width=80mm,height=50mm]{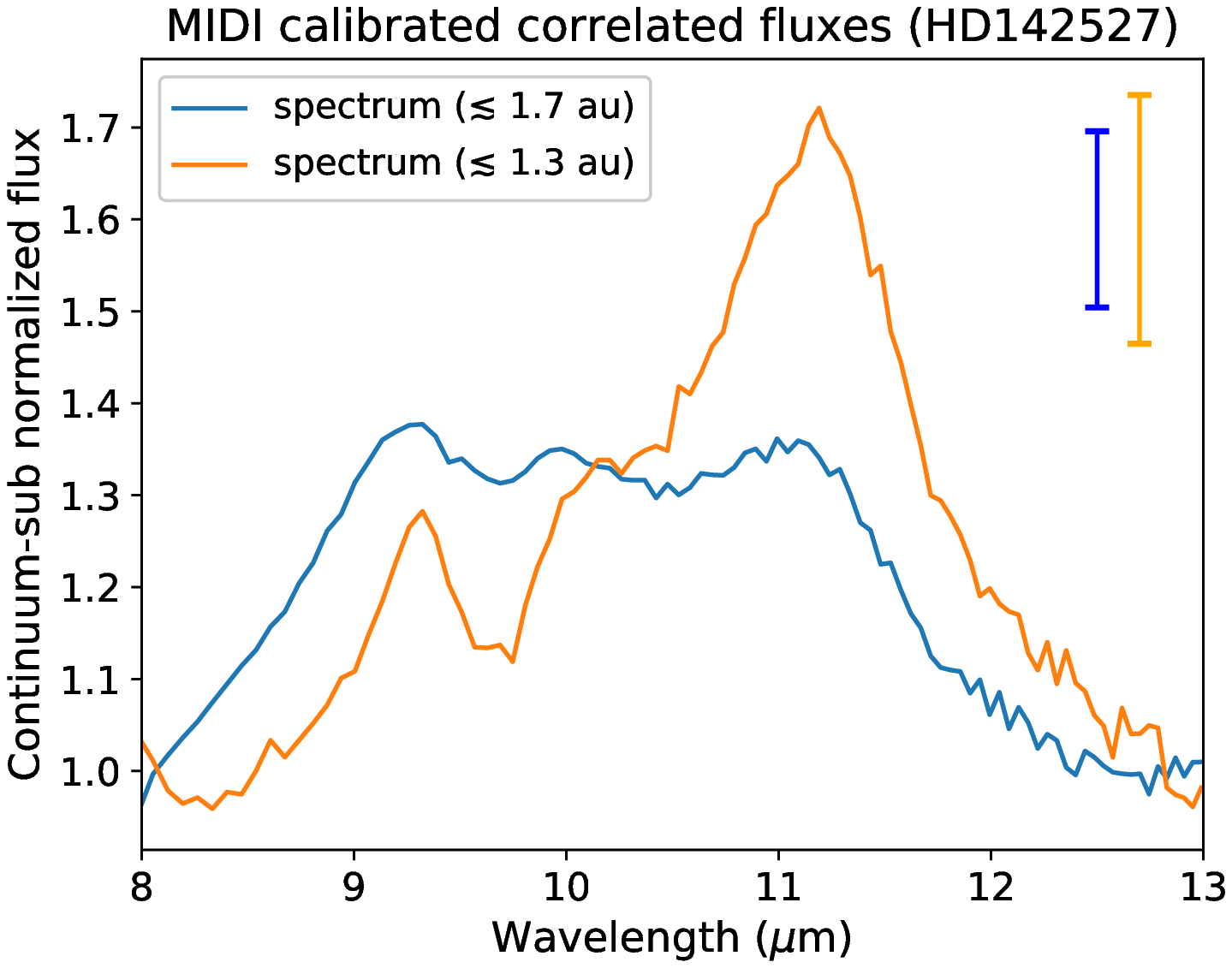}
 \includegraphics[width=80mm,height=50mm]{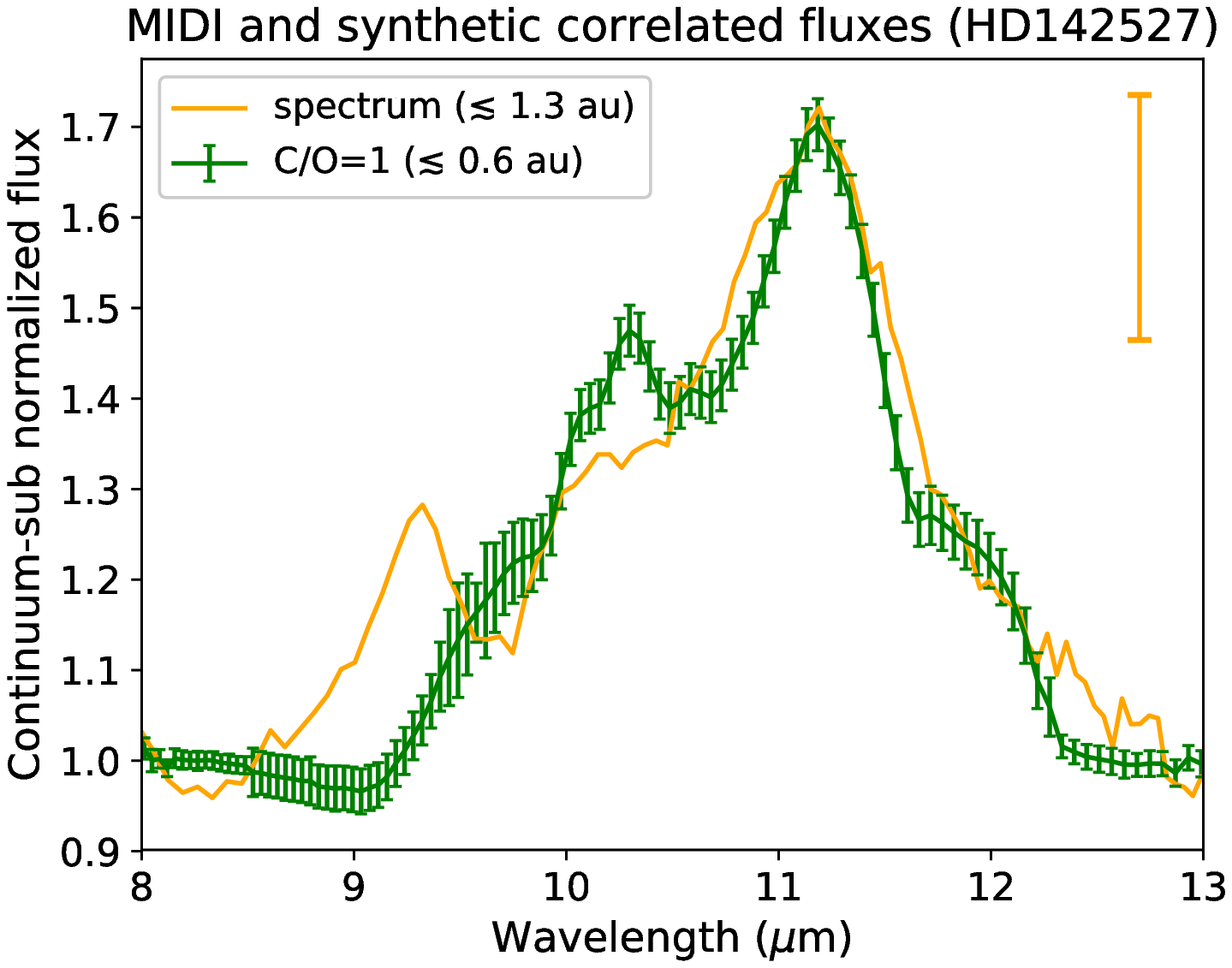}\\
 \includegraphics[width=80mm,height=50mm]{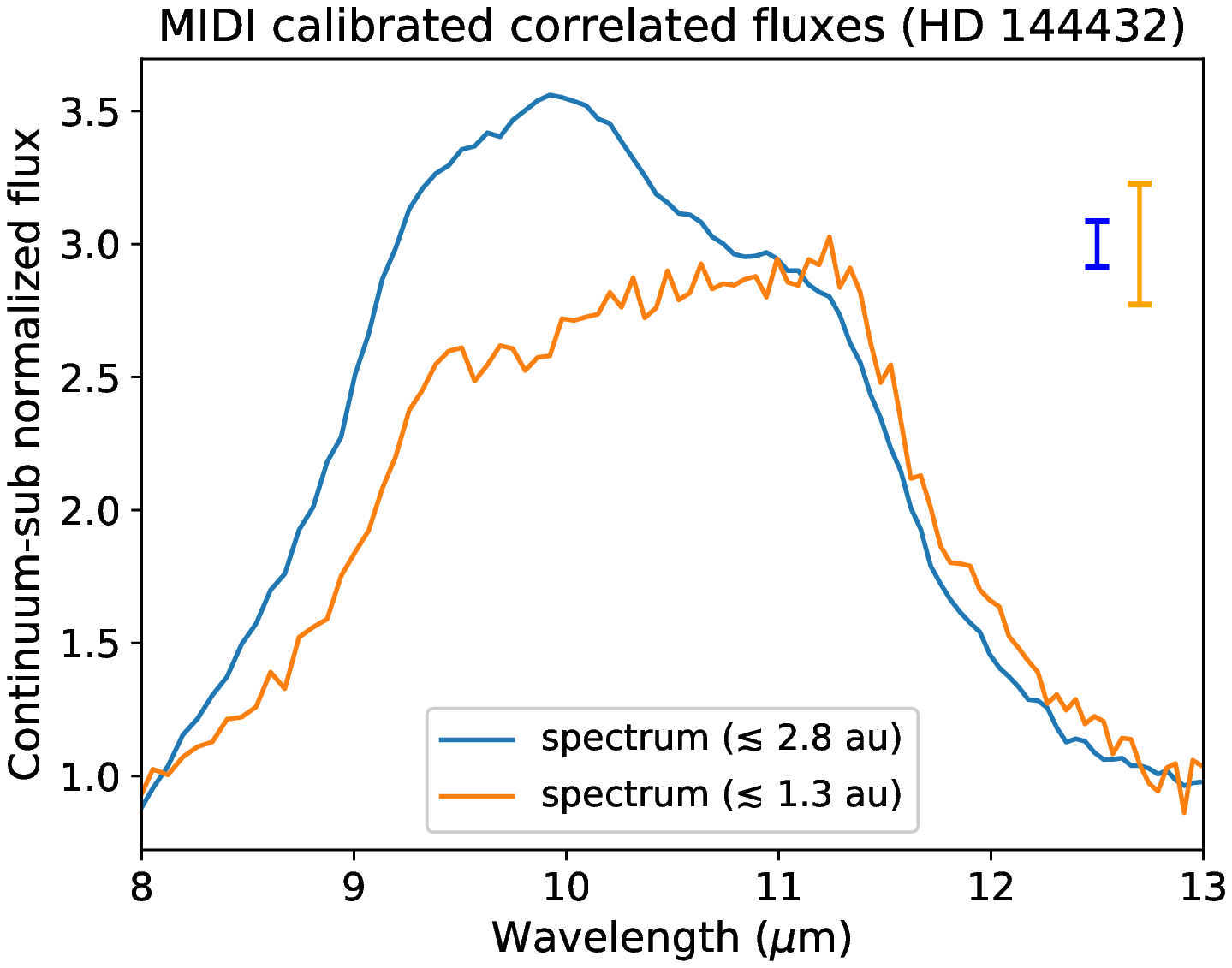}
  \includegraphics[width=80mm,height=50mm]{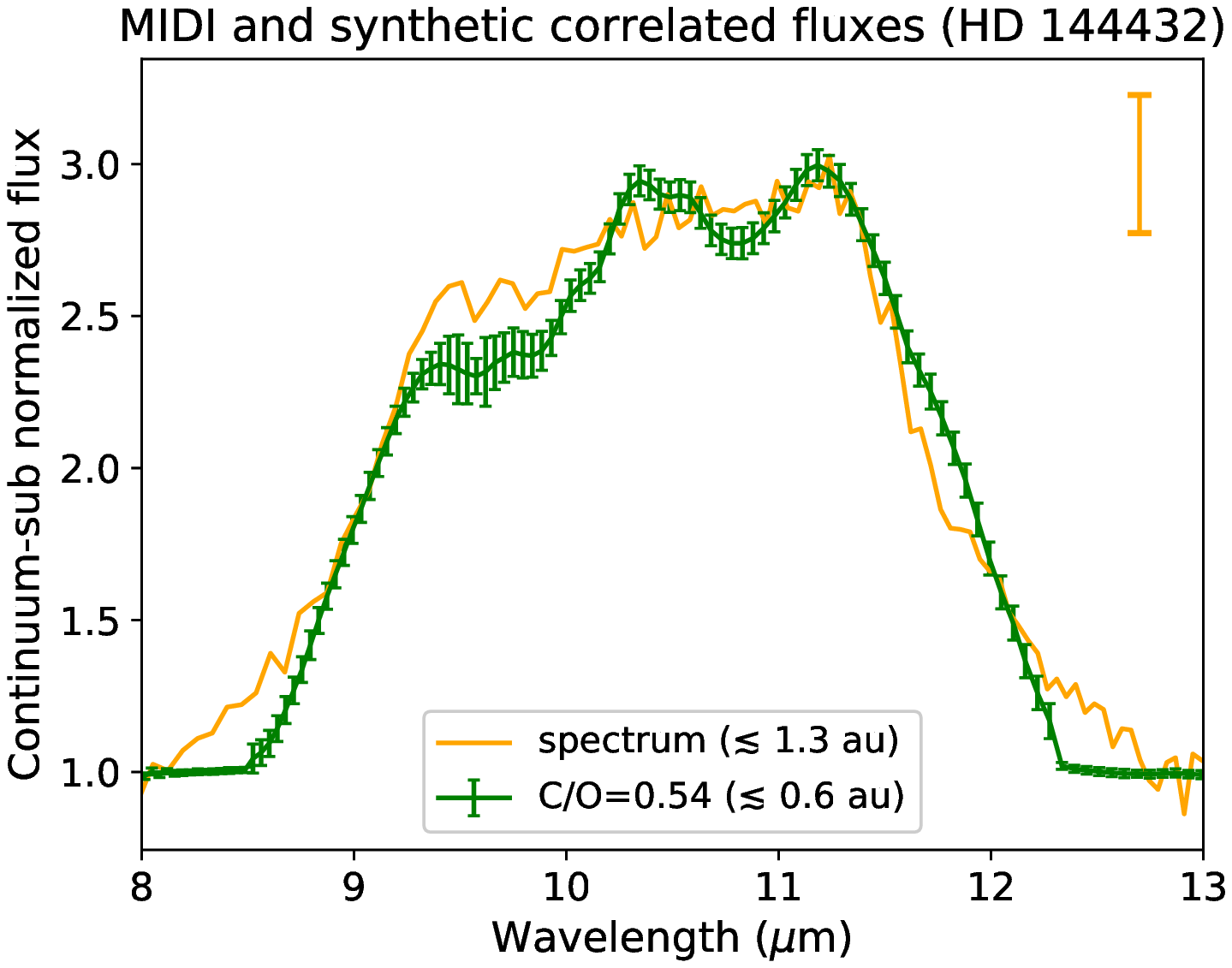}\\
 \includegraphics[width=80mm,height=50mm]{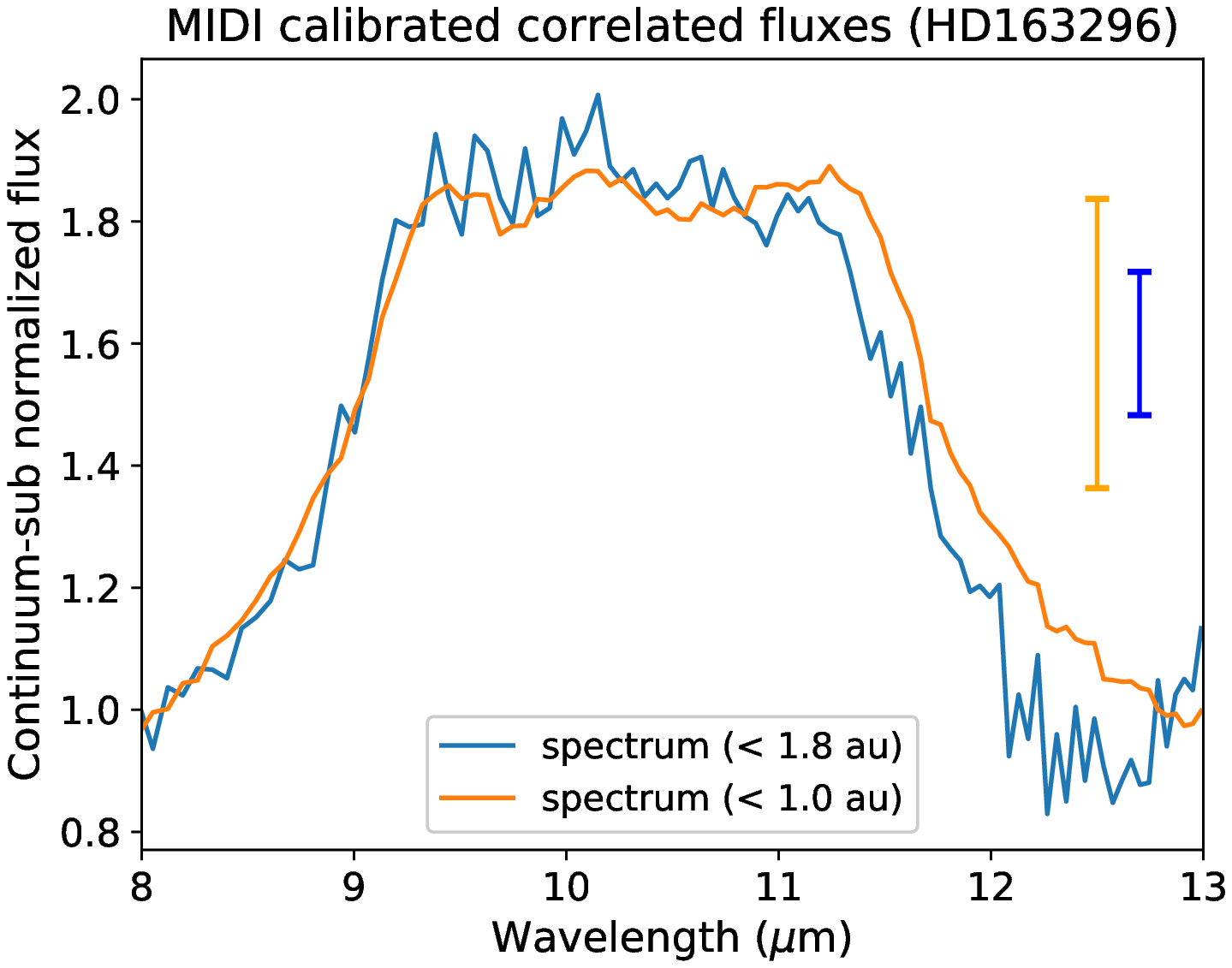}
 \includegraphics[width=80mm,height=50mm]{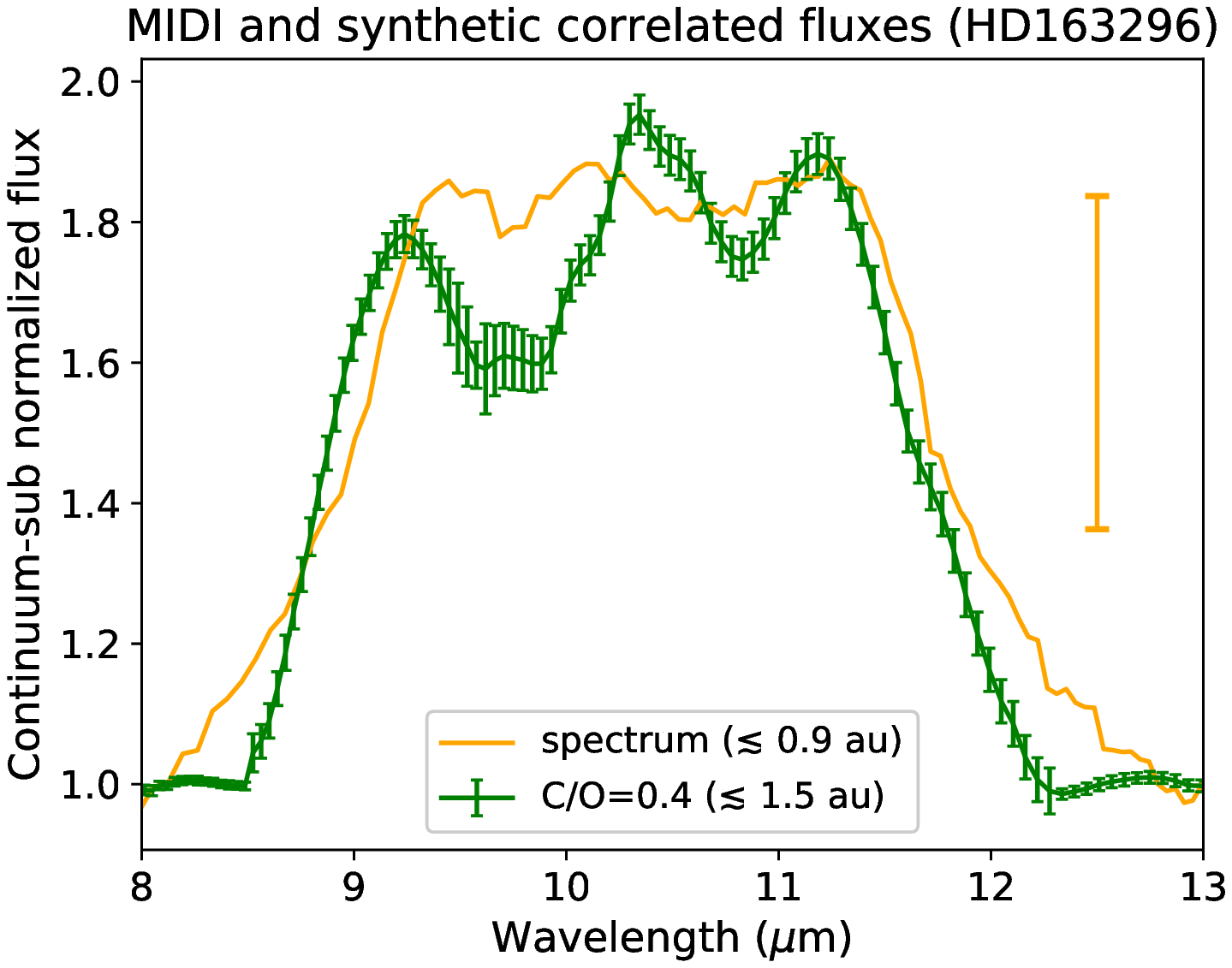}\\
 \includegraphics[width=80mm,height=50mm]{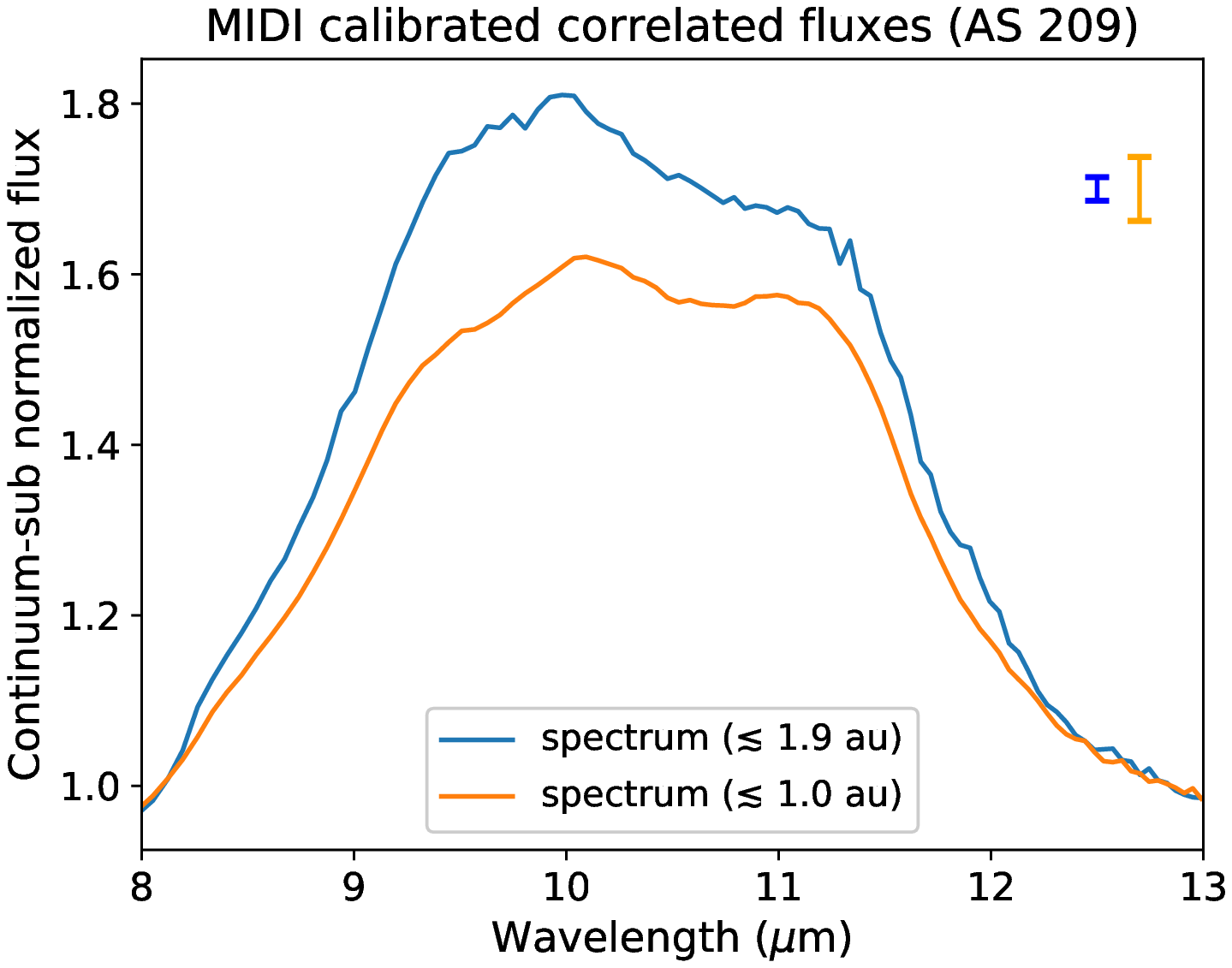}
 \includegraphics[width=80mm,height=50mm]{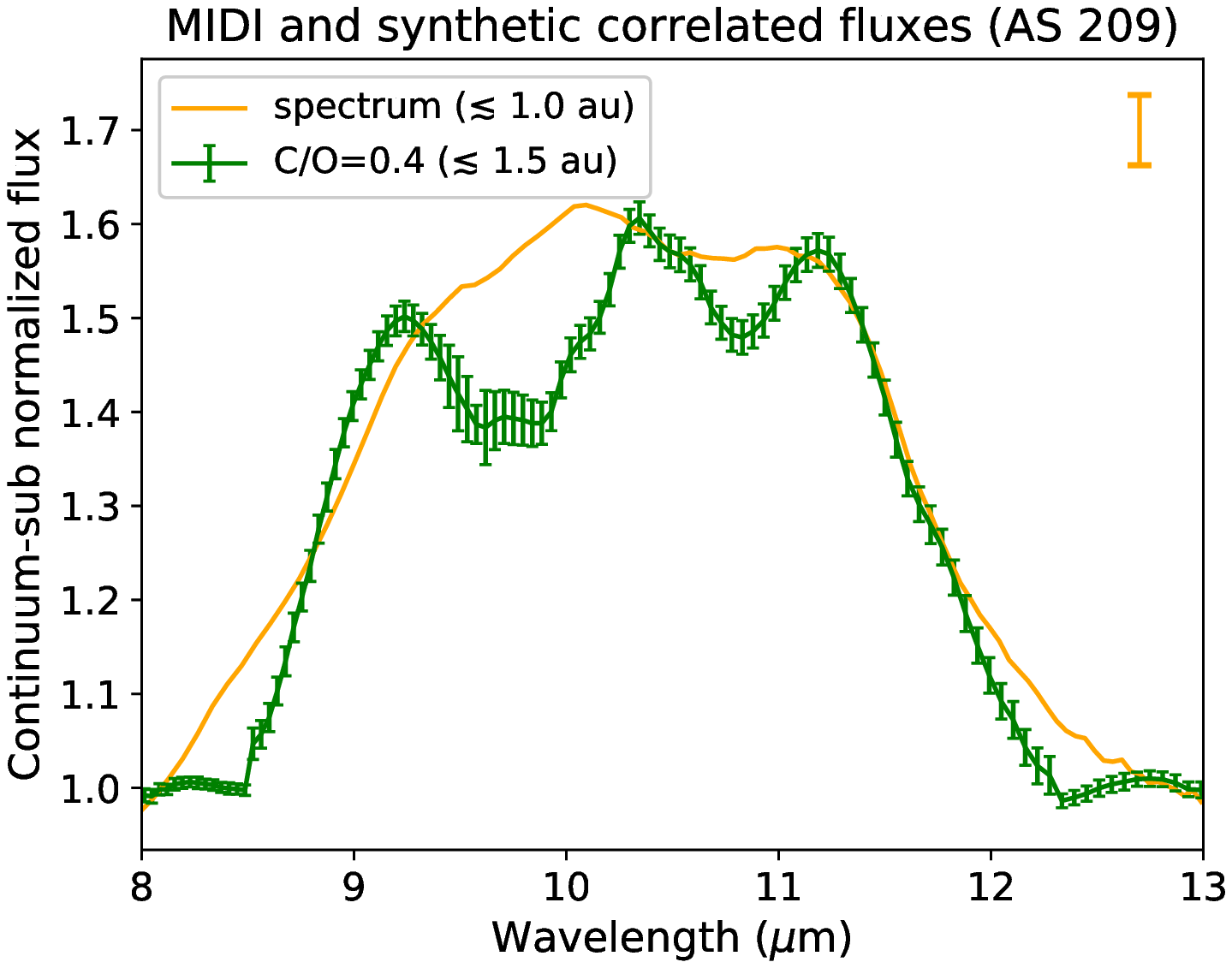}
  \caption{Left panels from top to bottom: Calibrated correlated spectra measured by MIDI for the three Herbig stars HD142527, HD144432, HD163296, and the T Tauri star AS209. The disc region probed by the different MIDI spectra is indicated in each panel; The uncertainties affecting the measured spectra are indicated by the single error bars in the top right corner of each plot. Right panels from top to bottom : innermost MIDI spectrum of the four sources overplotted with our innermost synthetic spectrum matching qualitatively the best the observed one; here, the high-resolution synthetic spectra were artificially scaled up to match the emission level of the MIDI spectra and smoothed to the MIDI spectral resolution (R=30).}
\label{fig:MIDI_res}
\end{figure*}

\subsection{Observational analysis}
 A first qualitative analysis can be made on the evolution of the observed MIDI spectra, shown in Fig.~\ref{fig:MIDI_res}, with respect to the probed inner disc regions. For HD142527, the change is drastic. The spectrum passes from a very flat emission band and without clear crystalline peaks to a red spectrum with a very prominent 11.3~$\mu$m forsterite peak. Another peak probably due to enstatite seems visible around 9~$\mu$m. As a caveat, there is a possibility that such drastic change could also be related to compositional changes either azimuthal and/or in time in the inner disc. The two spectra were indeed taken 9 years apart with different orientations of the baselines. In the case of HD144432, the more outer spectrum shows a triangular shape reminiscent of an amorphous interstellar silicate-based composition. However, as we enter the 2 au inner region, the spectrum get redder with more contribution from the 11~$\mu$m area. This is reminiscent of a significant dust processing and change in dust composition. For HD163296 the analysis is less clear as the error bars for the innermost spectrum are rather large. Nevertheless, when looking to the curve profile itself, the spectrum looks very flat, indicating large grains, and bluer than the spectrum from HD144432. Finally, for AS209, no clear change between the 2 and 1 au spectra is retrieved, with both suggesting a rather amorphous dust content. Nevertheless, we can note a decrease of the global amplitude of the emission band coupled with a slight increase of the forsterite peak at 11.3~$\mu$m. Both are reminiscent of stronger grain processing when moving toward the innermost disc region. Given the lower temperature and luminosity of AS209, it is very likely that these MIDI observations did not resolve enough the inner disc to see the predicted chemistry effects occurring at high temperature.

\subsection{Mineralogical interpretation}

{\bf comparison with synthetic spectra} Here, we aim to provide a first mineralogical interpretation of the MIDI spectra in the frame of our synthetic spectra. We thus performed a qualitative comparison between the observed and synthetic spectra to highlight possible similarities and trends. For that, we smoothed the high-resolution synthetic spectra ($R=220$) to the spectral resolution of MIDI ($R=30$) and scaled them up to match the global emission level of the measured MIDI spectra. We then selected the type of synthetic spectrum, characterised by a $C/O$ value and a minimum grain size , which decently reproduces the main spectral characteristics (features position and amplitude, spectral slope) identified in every observed innermost spectrum. They are all shown in the right panels of Fig.\ref{fig:MIDI_res}.\\
Following that qualitative approach, the innermost spectrum of HD142527, when probing the inner 1 au, presents an acceptable agreement with the synthetic $C/O=1$ spectrum with $a_{\rm min}=0.05$~$\mu$m. The very prominent 11.3~$\mu$m forsterite peak is well reproduced as well as the significant flux decrease outside the 11.3~$\mu$m feature. However we see an emission at 9.2 usually associated with enstatite (and marginally forsterite). This feature is not present in our synthetic $C/O=1$ spectra. As a consequence, for HD142527, we may be in presence of a solar $C/O$ ratio but with a subsolar $Mg/Si$ ratio. Indeed, subsolar values favour the stability of enstatite and silica while supersolar $Mg/Si$ values favour the stability of forsterite \citep{2001A&A...371..133F}. Moreover, condensation of forsterite can fractionate the gas in its magnesium and silicon content, which would thus reduce the $Mg/Si$ ratio and allow further condensation of enstatite- and silica-rich dust \citep{2016MNRAS.457.1359P}. Another possibility is that the solid components forming at $C/O=1$ and missing in the dust distribution we actually used (see Section~\ref{discchem}), would be responsible of the features in the HD142527's spectrum that are not present in our $C/O=1$ synthetic spectrum.\\ 
In the case of HD144432, a good agreement could be found with a solar synthetic spectrum ($C/0=0.54$) with $a_{\rm min}=0.05$~$\mu$m. Indeed, the increasing spectral profile as a function of wavelength, as well as a visible 11.3~$\mu$m forsterite peak, are both reproduced. Moreover, the flux decrease at the edges of the silicate band is also caught by our solar synthetic spectrum. Only the part of the observed spectrum around 9.5~$\mu$m, associated with enstatite, is slightly underestimated by our synthetic spectrum. However, such difference is within the estimated MIDI error bars, and is thus not fully significant.\\  
For the two other sources, HD163296 and AS209, no clear good match could be identified with the synthetic spectra. For HD163296, the comparison is made difficult due to the large error bars affecting the innermost MIDI spectrum. Nevertheless, a subsolar $C/O=0.4$ spectrum with $a_{\rm min}=0.05$ seems to reproduce some of the main characteristics of the observed spectrum, i.e. a bluer emission profile and a clear forsterite peak at 11.3~$\mu$m. However, the two spectra do match very well for the two apparent broad features around 9.5 um and around 10 um, probably related to enstatite again. Indeed, the bluest peak in the synthetic spectrum is mainly due to silica and peaks at about 9~$\mu$m, while the central feature is due to forsterite and thus peaks around 10.5~$\mu$m. Again, the synthetic spectrum does not seem to capture the enstatite part of the observed spectrum.
In the case of AS209, as mentioned previously, no clear change between the 2 and 1 au spectra is found. The innermost disc spectrum still looks "amorphous". A decent agreement could be found again with a subsolar synthetic spectrum in terms of emission level around 9~$\mu$m and around the forsterite peak at 11.3~$\mu$m (Forsterite). However, the synthetic spectrum significantly underestimates the flux between 9.5 and 10~$\mu$m, including the slight peak at $\sim$ 10~$\mu$m  (probably a mix between the close-by enstatite and Forsterite features). A lack of enstatite in our subsolar composition may thus again be at cause since it has a significant opacity contribution in that spectral window (see Fig.\ref{fig:opac}).

{\bf comparison with literature} Now, we want to check if the tentative mineralogy trends we derived on those objects are consistent with stellar abundance measurements, which represents the bulk composition of the central star photosphere, and with previous disc mineralogy measurements.\\
Regarding stellar abundances, \citet{2010IAUS..265..352C} derived atmospheric parameters and chemical abundances for a sample of Herbig stars using the VLT spectrograph FEROS. They measured subsolar $C/O$ values of 0.42 and 0.28 for HD144432 and HD163296, respectively.
More recently, \citet{2015A&A...582L..10K} collected stellar abundances of Herbig stars, which included HD144432 and HD163296. The values are the following :
\begin{itemize}
    \item HD144432 : $C/O=0.24\pm0.11$ and $Mg/Si=2.24\pm1.25$
    \item HD163296 : $C/O=0.31\pm0.21$ and $Mg/Si=1.48\pm0.65$
\end{itemize}
According to stellar abundance measurements, both stars seem to show a subsolar $C/O$ composition ($C/O < 0.54$) and a supersolar $Mg/Si$ composition ($Mg/Si > 1.23$), even though the large error bars affecting the $Mg/Si$ measurements makes the comparison not conclusive. Nevertheless, the subsolar $C/O$ trend we derived for the HD163296's disc, from our model spectra, appears consistent with the bulk composition of the host star. However, that is not the case for HD144432 for which we favoured a solar $C/O$ composition for the disc. Two possible explanations for that apparent discrepancy could be : 1) the stellar photosphere is carbon-depleted with respect to the disc, and/or 2) the $C/O$ and $Mg/Si$ ratios are indeed respectively subsolar and supersolar, which would imply a higher amount of forsterite in the disc (compared to the case of solar $C/O$ and $Mg/Si$). In the latter case, the resulting 'forsterite-rich' synthetic spectrum could thus mimic the purely  solar synthetic spectrum (solar $C/0$ and solar $Mg/Si$) we already produced.

Regarding the discs mineralogy, a first comparison can be made on the basis of SPITZER data. \citet{2010ApJ...721..431J} made a compositional analysis of the N-band solid state features on a significant number of Herbig sources. The dust composition they derived for our three sources of interest are :

{\bf HD144432}:  48\% amorphous olivine, 34\% amorphous pyroxene, 7\% silica, 5\% crystalline enstatite. 

{\bf HD142527}: 50\% amorphous olivine, 22\% amorphous pyroxene, 5.7\% forsterite, 6\% enstatite, 10.5\% silica.

{\bf HD163296}: 62\% amorphous olivine, 18\% amorphous pyroxene, 8.5\% silica, 2.6\% forsterite, 4.5\% enstatite.




Those compositions were derived from spatially unresolved observations which include both the inner and outer regions. Our calculations (see table 1) show that the mineralogy of the inner disc can be very specific. Most of the dust would be processed and thus, condensed or annealed into crystalline form, while the presence of amorphous phases is unlikely. Another important difference is the high amount of metallic Fe, present in our calculations, and which is highly opaque in the infrared. Our derived compositions for the inner disc, clearly show differences. As a consequence, there is the need to update the fiducial dust composition when modelling the emission spectra of resolved discs. \\
Another comparison can be made with the study of \citet{2004Natur.432..479V}, which focused on the inner disc region (R$\lesssim$3~au) composition, especially on the crystalline forsterite/enstatite ratio. For all the considered sources, they derived forsterite-rich dust with fo/en$\sim$2. In our case, when averaging the high and low temperature region according to the spatial resolution of \citet{2004Natur.432..479V}, we predict a fo/en$\sim$1.1 for the solar case, fo/en$\sim$3.8 for the C-rich case, and fo/en$\sim$0.86 for the subsolar one. Moreover, a fo/silica$\sim$3.15 is predicted for a subsolar composition. Our identified trends for the inner disc of HD144432 and HD142527, i.e. solar and supersolar $C/O$ ratio, agrees with the forsterite rich dust (fo/en$\gtrsim$2) derived by \citet{2004Natur.432..479V}. However, while we identified a subsolar composition trend for HD163296, \citet{2004Natur.432..479V} still derived a forsterite-rich dust (fo/en$\gtrsim$2). Interestingly, the significant emission around 9 micron, which does not belong to forsterite, was not captured by the compositional fit of \citet{2004Natur.432..479V}, possibly indicating a change in the dust chemistry.

\section{Discussion}
\label{discussion}

{\bf Current dust mixture}. Our results show that fine tuning the chemistry is necessary if we want to interpret spatially resolved N-band spectra of discs. While a single general dust mixture can be appropriate for spatially unresolved observations, it can fail when trying to model the resolved N-band spectra of the inner discs region, for instance the inner 1-2 au. In fact, given the high temperatures observed in these disc regions, the radial change of the dust chemistry that results from the condensation and annealing of the different material, would have a strong impact in the resulting spectra. The resulting dust mixture would not likely match the composition of the dust reservoir in the cooler regions of the disc. Different dust mixtures for different resolved spatial scales can return better fitting. 
However, it is important to check the consistency of the list of species that is used. The dust species selected for fitting the disc spectra of the inner hotter disc region have to be consistent with the  chemical elemental composition of the bulk disc. For example, if solar-like composition is used, the high temperature region do not see the coexistence of silicates and solid carbon material. Moreover, silica is not a stable phase for solar $C/O$, unless gas-dust fractionation occur or there are strong changes in the $Mg/Si$ ratio. On the other hand, in carbon-poor environment, a series of oxides has to be considered in the list, while for $C/O\ge1$, several carbon species replace oxides (see section~\ref{discchem}). Avoiding inconsistency is, thus, extremely important when choosing the fiducial dust mixture.\\

{\bf Source of opacity}. It is commonly assumed that graphite and amorphous carbon are among the most efficient carriers of infrared opacity. These compounds have been used as fiducial carrier for most of the  dust mixtures used for spectra modelling. Carbon-rich dust can be present in the outer disc regions, as pre-stellar grains or trapped in ice particles. However, considering carbonaceous dust species as the main source of opacity in the inner disc regions is likely incorrect for solar and subsolar compositions. Our fiducial dust compositions for the inner disc have large amount of iron for solar and $C/O=1$, and oxides for subsolar $C/O$. Iron, FeO and MgO compounds are basically featureless in the mid-infrared, but still strong contributors to the dust continuum. On the other hand, solar and subsolar compositions do not have any carbon solids. Therefore, we suggest that carbon may be replaced by iron when modelling the inner disc spectra. In particular, having the correct solid species contribution to the disc opacity is key for a proper interpretation of the spatially resolved observations of the inner dust rim of discs, and thus the nature of the innermost hot dust. Both the dust composition (through the sublimation temperature) and grain size (through the cooling efficiency and scattering anisotropy) have a direct effect on the location and appearance of the dust inner rim \citep[see e.g.,][]{2018A&A...609A..45C}. That can lead to a degeneracy in the modeling since a same inner dust rim profile and location can be reproduced with different pairs of dust composition/grain size.\\  

{\bf Dichotomy}. The  differences in the spectra produced when varying the $C/O$ ratio, would allow, if confirmed by observation, to disentangle different bulk disc compositions. This could help to reconcile with the apparently broad range of $C/O$ compositions observed in planet-hosting stars (see section\ref{intro}). 
However, if no significant difference is to be found between the resolved spectra of the inner region of discs, Other possibilities will have to be considered:
\begin{itemize}
\item The chemistry in the disc surface is largely and extremely different from the midplane  \citep{2015A&A...582L..10K}. This can be due to the large variation in temperature and pressure that different disc regions experience. 2D disc models show that the discs midplane, because of lower temperatures \citep{Dalessio1998} can contain large amount of unprocessed dust \citep{2016MNRAS.457.1359P}. As a consequence, what we observe would not correlate with the chemistry and dynamics of the dust and gas that characterize the disc and may have fed the forming star.
\item There would actually be no significant variation in the $C/O$ ratio of star forming regions. 
Such possibility is suggested by galactic chemical evolution models, which predict that the $C/O$ ratio should peak around the solar value without much variation \citep[e.g., ][]{2015ApJ...804...40G}. It would mean that the range of $C/O$ ratios found in nearby stars (0.2<C/0<0.8) is actually narrower, which could be related to the difficulty in measuring the atomic C and O lines. Indeed, 
\citep{2012ApJ...747L..27F}, and more recently \citet{2016ApJ...831...20B}, suggested that the high $C/O$ values derived for F,G and K stars might be overestimated. The reason lies in the difficulty in determining the abundances of carbon and oxygen in the observed stellar photospheres which might depend on the spectral lines and the method used for the lines analysis \citep{2012ApJ...747L..27F}.


\item It is now widely accepted that a strong connection between the star, the disc and the parent cloud was in place at early stages. Therefore, the chemistry of the star and disc was thus strictly related to that of the cloud and to the processes that transferred the material. Simulations show that dust and gas that are injected from the cloud into the forming disc at different times and locations can indeed experience widely different thermal conditions  \citep{2005A&A...442..703H,2012M&PS...47...99Y,2018ApJ...867L..23P}. As such, the material that was incorporated by the star during the cloud collapse could have been of different nature and composition (heterogeneous in C and O for example) than the portion of the cloud that fed the disc. 

\item we would not actually be resolving enough the inner regions where the predicted thermodynamics condensation reactions occur. The presence of unprocessed dust from the outer regions, which would be amplified by a possible strong radial mixing, would erase the features of the predicted chemistry.  

As an illustration of possible intrinsic heterogeneities between the star and the different disc regions, the case of HD142527 is particularly interesting. The disc around HD142527 presents a complex structure made of a warped inner dust disc and an outer disc, which are separated by a large 100 au-wide gap. A low-mass companion was also found orbiting the central star at about 11 au, just outside the inner disc \citep{2018A&A...617A..37C,2019A&A...622A..96C}. Additional large-scale structures like holes and spirals were also detected \citep{2014ApJ...781...87A}. By comparing the observed MIDI spectra with our synthetic spectra, we found a very good agreement with a dust composition that is enriched in C. On the other hand, IR scattered-light spectra of HD142527 showed a significant \ce{H2O}(ice) absorption feature \citep{2009ApJ...690L.110H}. Both findings are in theory inconsistent since high $C/O$ ratio would predict a lack of oxides including water. To reconcile observation and modelling, as previously stated, different explanations are possible: (i) HD142527 could actually be characterised by different bulk chemical reservoirs that fed the star and the different disc regions, or (ii) the redistribution of oxygen between \ce{CO} and \ce{H2O}, which occurs at lower temperatures ($T\leq700$~K) and produces secondary compounds such as methane, as predicted by equilibrium calculations  \citep{1975GeCoA..39..389L,1999IAUS..191..279L,2018A&A...614A...1W}, is efficient.
\end{itemize}

{\bf Parallel observations} To better understand the chemical and physical conditions within the discs, observations of other gas, solids or ice components in different spectral bands may provide insightful information. For example, the presence of methane or other sub-products (if detectable) observed in spectra of high $C/O$ candidates, as in the case of HD142527, would explain the presence of water. On the other hand low $C/O$ ratios would return a large amount of water, and different oxides as pointed out in the first part of this work. Carbon-rich molecules like PAHs could also trace the $C/O$ chemistry by showing differences in their associated IR features, which were observed around many Herbig stars and some T Tauri stars \citep{2017ApJ...835..291S}. 
Therefore, we can see the importance of companion compounds such as methane in establishing the mechanisms at the basis of the discs solid chemistry. In that context, a systematic study of the optical properties and detectability of the solids, gas and ices, predicted at different $C/O$ ratios, in their respective wavelength domain is essential. \\
\\

{\bf Impact on grain growth and planet formation}. The distribution of solids and their dynamics in a disc drives their growth up to the formation of planetary embryos (rocky planets) and cores of giant planets. In that context, the various condensation fronts of refractory and volatile elements (such as water or CO) can be important since they set the different locations beyond which the density of solids can significantly increase \citep[see e.g.,][for a review]{2014prpl.conf..363P}. In our Solar System, such density increase beyond the water ice line (the so-called snow-line) is commonly thought to have segregated the formation of small rocky planets (inside) and giant gaseous planets (outside). Actually, the associated density increase seems to have been unsufficient to explain fully that dichotomy especially against other processes such as pebble accretion \citep{2015Icar..258..418M}. Nevertheless, it remains an important element that drove the distribution of solids of different compositions in the early Solar System, and thus the solid composition of the planets \citep{2016Icar..267..368M}.\\ 
In a given protoplanetary system, the existence of specific condensation fronts and their impact in terms of solid density increase is directly related to the bulk composition of the protostellar nebula and the associated condensation sequence. For instance, the widely reduced amount of water that is associated with high $C/O$ ratio could erase the impact the snow-line is expected to have on planet formation and composition. The separation between 'volatile-poor' and 'volatile-rich' planets and minor bodies would then occur around the ice line of other more abundant volatiles like CO, but, in this case, at very low temperatures. On the other hand, a subsolar $C/O$ ratio would enhance the formation of giant planets at the snow-line (and beyond) given the predicted massive amount of condensed water ice. The solid composition of the formed planets would then be driven by these differences and amplified by the possible fossilization of those condensation fronts, as it may have happened in our Solar System \citep{2016Icar..267..368M}.\\ At early stages, during the disc phase, these differences may also impact the dust grain properties. For instance, the lack of water ice induced by a supersolar $C/O$ initial bulk composition would imply less icy and thus less sticky cold grains, which decreases the grain growth efficiency \citep{2018SSRv..214...52B}. Less efficient grain growth over a large portion of the disc could prevent the formation of dusty structures (rings, crescents, clumps) observed in many discs \citep{2018ApJ...869L..41A}, and commonly thought to be due to the local accumulation of partially decoupled large dust grains ($\gtrsim 1$~mm) in gas pressure bumps \citep[e.g.,][]{}. The absence of marked features in the mm-sized dust distribution of a disc could thus be another clue of its volatile-poor content.\\

{\bf Limitations and expected improvements} Our produced synthetic spectra were derived from a static 1D disc at thermodynamic equilibrium. The aim of our work was indeed to study the impact of a fundamental chemistry parameter, the bulk $C/O$ ratio, on the radial variation of dust composition and the related mid-infrared solid-state features, under given fixed conditions. Therefore, our work did not include different disc models nor the modelling of specific objects (different central star properties, discs gaps and cavities, disc densities, dust properties such as crystalline fraction and transport), that are, indeed, required when modelling the observed spectra. 
More realistic calculations should consider a full 2D disc model with a composition that would vary both radially and vertically in the optically thin zones. In addition, an iterative process should also ideally be carried out between the radiative transfer disc model, thermodynamic calculations and kinetics consideration. Eventually, dynamics of discs (e.g. different growing/fragmentation and dust settling regimes) should also be included.
Moreover, as mentioned in Section \ref{observation}, variation in the abundances of other elements can affect the dust composition as these can have, of course, different abundances from the Sun. For example, $Mg/Si$ ratio can also have an impact on the resulting dust composition. Such an exploration of the $Mg/Si$ parameter is foreseen. Another limitation of our modelling work is the lack of optical constants in the literature for some of the dust components that are stable at the different $C/O$ ratios we explored. Therefore, we simplified our obtained dust mixtures accordingly. Future optical constant measurements on the missing dust components will be important for more detailed modelling of the dust opacities used for radiative transfer modelling of inner-disc spectra.   

\section{Conclusions}
\label{conclusions}
 The advent of new MIR spectroscopic instruments like MATISSE, and soon the JWST/MIRI \citep{2015PASP..127..584R} and E-ELT/METIS \citep{2018SPIE10702E..1UB} instruments, will allow to observe and study in detail the solid chemistry of discs down to their innermost regions ($\sim 0.1$ to 10 au). In this context, a fine theoretical basis and description of the chemistry in the inner disc regions is essential. 
 
 In this work we derived synthetic infrared spectra of a fiducial protoplanetary disc model with different dust compositions for the inner disc regions ($r<1$~au). Those compositions were derived from condensation sequences computed at LTE for different initial bulk $C/O$ ratios (subsolar: $C/O=0.4$, solar: $C/O=0.54$, and supersolar: $C/0=1$), which is a fundamental parameter of the solid chemistry 
Our aim was to analyse the effect, on the observables, of different inner disc dust compositions in order to determine to which extent the initial bulk composition of discs and the associated condensation sequence can be constrained from the observations. Moreover such a work constituted also a first effort to provide reliable dust mixtures for the modelling and interpretation of future spatially resolved observations of the inner disc regions. 

The main results of our work are the following :
\begin{itemize}
    \item Similarly to previous works on condensations sequences, the condensation sequences associated with subsolar and supersolar $C/O$ ratios produce rather different dust compositions in the inner 1 au, compared to the solar case which produces mostly silicate (forsterite and enstatite). In the subsolar case, more than 60\% of the condensates are in the form of oxides (Iron oxide, magnesium oxide, silica). In the supersolar case, most condensates are under the form of forsterite and iron-rich species (FeSi), and even almost exclusively FeSi in the high temperature (T>1000K) region.
    \item The three different dust compositions return very different synthetic MIR spectra, especially when considering that small sub-micron-sized dust grains are present. The spatially resolved (or inner-disc) spectra present a rather flat profile with a prominent silica/enstatite peak around 9~$\mu$m and forsterite peak at 11.3~$\mu$m in the $C/0=0.4$ case, a slightly redder profile with prominent forsterite peaks around 10~$\mu$m and 11.3~$\mu$m in the $C/0=0.54$ case, and a very red profile with a very prominent forsterite feature at 11.3~$\mu$m and almost no emission shorter than 10~ $\mu$m in the $C/0=1$ case. The amplitude of the N-band emission band is the greatest in the $C/0=0.54$ case.
    \item Our analysis of the synthetic inner-disc spectra suggests that MATISSE should be able: 1) to detect the differences between different $C/0$ initial bulk compositions and 2) to trace, down to the sub-au scale, the relative radial change in chemistry associated with a given bulk composition.
    \item We used our synthetic spectra to perform a first interpretation of existing `inner-disc' spectra obtained with the former VLTI instrument MIDI, on three Herbig stars (HD142527, HD1444432, HD163296) and one T Tauri star AS209. Two objects can be directly associated with a specific $C/O$ bulk composition: HD142527 for the 'carbon-rich' (supersolar) bulk composition and HD1444432 for the more 'oxygen-rich' (solar) one. The stellar photosphere compositions appear consistent with our interpretation. The two other objects does not show a very good agreement, which suggests more complexity in the dust chemistry and/or disc properties that are not captured by our synthetic spectra. Moreover, in the case of the colder and less luminous T Tauri star AS209, the MIDI observations did not probably resolve enough the inner disc to see the predicted chemistry effects occurring at high temperature.
    \item Comparison of our results with the literature shows that the inner disc mineralogy can be very specific and not related to the global dust compositions derived from spatially unresolved observations like SPITZER observations. 
\end{itemize}
Our work highlights the need for including more complex dust chemistries in the modelling and interpretation of solid-state spectroscopic observations targeting especially the inner regions of discs. Such more complex chemistries will require optical constant laboratory measurements on a greater number of solid species. As a perspective of our work, incorporating dynamical and chemical modelling of the disc should be pursued to include the effect of dust growth, fragmentation, drift and mixing.   

\section*{Acknowledgments}
The authors wish to thank the anonymous referee for their useful comments that let us investigate in more detail the limitations of our proposed model and the interpretation of results. The authors wish to thank Matthieu Gounelle for valuable discussions. This research has made use of the Jean-Marie Mariotti Center \texttt{OiDB} service \footnote{Available at http://oidb.jmmc.fr}. F. C. Pignatale wishes to acknowledge the financial support of ANR-15-CE31-0004-1 (ANR CRADLE) and of the R\'{e}gion \^{I}le-de-France through the DIM-ACAV + project: ``HOC - Origine de l'eau et du carbone dans le Système Solaire''.

\section*{Data availability}
All the raw and calibrated observational data underlying this work are publicly available at http://archive.eso.org and at http://oidb.jmmc.fr, respectively.

\bibliographystyle{mnras}
\bibliography{biblio} 








\bsp	
\label{lastpage}
\end{document}